\documentclass[11pt,a4paper]{article}
\usepackage{jheppub}
\usepackage{graphicx}
\usepackage{rotating}
\usepackage{pdflscape}

\def\be{\begin{equation}}
\def\ee{\end{equation}}

\newcommand\MSbar{$\overline{\mathrm{MS}}~$}
\newcommand\as{\alpha_s}
\newcommand\aS{\alpha_s}
\newcommand\mcO{\mathcal{O}}

\newcommand\muzf{\mu_{\scriptscriptstyle\rm 0F}}
\newcommand\muzr{\mu_{\scriptscriptstyle\rm 0R}}
\newcommand\muf{\mu_{\scriptscriptstyle\rm F}}
\newcommand\mur{\mu_{\scriptscriptstyle\rm R}}

\preprint{{\raggedleft%
ZU-TH 79/23
}}

\title{Heavy Quark Fragmentation in \boldmath{$e^+e^-$} Collisions \\ to NNLO+NNLL Accuracy in Perturbative QCD}

\author[a]{Leonardo Bonino,}
\affiliation[a]{Physik-Institut, Universit\"at Z\"urich, 
Winterthurerstrasse 190, CH-8057 Z\"urich, Switzerland}

\author[b,c]{Matteo Cacciari,}
\affiliation[b]{Sorbonne Universit\'e, CNRS, Laboratoire de Physique 
Th\'orique et Hautes \'Energies,\\ LPTHE, F-75005 Paris, France}
\affiliation[c]{Universit\'e Paris Cit\'e, LPTHE, F-75006 Paris, France}

\author[a,d]{Giovanni Stagnitto}
\affiliation[d]{Universit\`{a} degli Studi di Milano-Bicocca \& INFN,
  Piazza della Scienza 3, Milano 20126, Italy}

\emailAdd{leonardo.bonino@physik.uzh.ch}
\emailAdd{matteo.cacciari@lpthe.jussieu.fr}
\emailAdd{giovanni.stagnitto@unimib.it}

\abstract{Fragmentation of heavy quarks into heavy-flavoured hadrons receives both perturbative and non-perturbative contributions. We consider perturbative QCD corrections to heavy quark production in $e^+e^-$ collisions to next-to-next-to-leading order accuracy in QCD with next-to-next-to-leading-logarithmic resummation of quasi-collinear and soft emissions. 
We study multiple matching schemes, and multiple regularisations of the soft resummation, and observe a significant dependence of the perturbative results on these ingredients, suggesting that NNLO+NNLL perturbative accuracy may not lead to real gains unless the interface with non-perturbative physics is properly analysed.
We confirm previous evidence that $D^{*+}$ experimental data from CLEO/BELLE and from LEP are not reconcilable with perturbative predictions employing standard DGLAP evolution. 
We extract non-perturbative contributions from $e^+e^-$ experimental data for  both $D$ and $B$ meson fragmentation. Such contributions can be used to predict heavy-quark fragmentation in other processes, e.g.\ DIS and proton-proton collisions.}

\keywords{Higher Order Perturbative Calculations, Resummation}

\begin{document}
\maketitle

\section{Introduction}

Fragmentation of heavy quarks, namely charm and bottom, into heavy-flavoured hadrons has been extensively studied in the past. Just a few years after the charm quark discovery the fragmentation function of heavy quarks was understood to be much harder than that of light quarks and peaked at large hadron energy fractions~\cite{Suzuki:1977km,Bjorken:1977md}, while  phenomenological models trying to describe its functional form soon appeared (see e.g.\ \cite{Kinoshita:1981af,Kinoshita:1985mh,Peterson:1982ak}). 

Perturbative calculations of heavy quark fragmentation were also soon considered, and resummation of collinear logarithms up to next-to-leading order accuracy, as well as of soft logarithms to leading logarithmic accuracy, was performed in~\cite{Mele:1990yq,Mele:1990cw}. Refs.~\cite{Dokshitzer:1995ev,Cacciari:2001cw} extended soft resummation to next-to-leading logarithmic accuracy, while~\cite{Cacciari:2005uk} performed a comprehensive phenomenological analysis at NLO+NLL level. The interplay between a massive quark and the resummation of soft gluons has been investigated e.g.\ in~\cite{Aglietti:2007bp,Aglietti:2022rcm,Gaggero:2022hmv,Ghira:2023bxr}.

A number of papers considered the interplay of perturbative and non-perturbative physics in this process using an array of techniques: heavy quark effective theory~\cite{Jaffe:1993ie,Neubert:2007je}, renormalons~\cite{Gardi:2003ar,Gardi:2005yi}, effective strong coupling~\cite{Dokshitzer:1995ev,Aglietti:2006yf,Corcella:2007tg}, soft and collinear effective theory and boosted heavy quark effective theory~\cite{Fickinger:2016rfd}.

Despite fixed order calculations to next-to-next-to-leading-order accuracy having been available since quite some time~\cite{Rijken:1996vr,Rijken:1996ns,Mitov:2006wy,Blumlein:2006rr,Melnikov:2004bm,Mitov:2004du}, it is only quite recently that they were implemented into codes and tools useful for studying them numerically and for doing heavy quark fragmentation phenomenology. Notably, bottom quark fragmentation at next-to-next-to-leading order (NNLO) plus next-to-next-to-leading logarithmic (NNLL) collinear and soft accuracy has been considered in~\cite{Fickinger:2016rfd,Ridolfi:2019bch,Maltoni:2022bpy,Czakon:2022pyz}, but the implications of the higher order perturbative accuracy have probably not been fully explored, and charm quark fragmentation has not been studied.

The purpose of this paper, which considers both charm and bottom fragmentation, and also addresses top fragmentation at very large energies, is to provide a phenomenological and user-friendly implementation of heavy quark fragmentation in $e^+e^-$ collisions at NNLO+NNLL level. 
The paper also aims to study more comprehensively the impact of NNLO+NNLL corrections, and their interface with the non-perturbative phase describing the hadronisation of the heavy quark into the heavy flavoured hadron. 
Finally, by comparing to experimental data the paper extracts a non-perturbative component which can be reused to calculate heavy hadron production in hadronic collisions.

Section~\ref{sec:setup} sketches the setup and establishes nomenclature and notations. Section~\ref{sec:numres} presents numerical results, while section~\ref{sec:fits} introduces comparisons to experimental data and extracts non perturbative components by comparing to them. All details and formulas, with the exception of the NNLO fixed order results of~\cite{Rijken:1996vr,Mitov:2006wy,Blumlein:2006rr,Melnikov:2004bm,Mitov:2004du}, too lengthy to be reproduced here, are given in a series of Appendices.

\section{Perturbative Setup}
\label{sec:setup}

We denote the cross section for the full, observable-level process for the production of a heavy-flavoured hadron $H$ in $e^+e^-$ collisions with an energy fraction $x = E_H/E_\mathrm{beam}$, i.e.\ a one-particle inclusive fragmentation function (FF), as
\be
\frac{d\sigma_H}{dx}(x, Q) \, ,
\ee
where $Q = 2E_\mathrm{beam}$ is the total-centre-of-mass energy of the collision, here implicitly understood to be that of electron-positron beams\footnote{We will use the same symbol $Q$ for the centre-of-mass energy of the $e^+e^-$ collision and for the heavy quark. We trust that this will not lead to confusion.}. This cross section will be modeled by splitting it into a convolution\footnote{We denote the convolution operation as $[f\otimes g](x) \equiv \int_x^1 dz/z f(x/z) g(z)$.} of a perturbative and a non-perturbative component, 
\be
\label{eq:ff}
\frac{d\sigma_H}{dx}(x, Q) \simeq 
\frac{d\sigma_Q}{dx}(x,Q,m) \otimes
D^{np}_{Q\to H}(x,\{params\}) \, .
\ee
The non-perturbative component, $D^{np}_{Q\to H}(x,\{params\})$, is here understood to  integrate to  $f_{Q\to H}$, the fraction of heavy quarks $Q$ hadronising into the given hadron $H$, whereas the perturbative component $d\sigma_Q/dx(x)$ is the cross section for producing the heavy quark $Q$ with energy fraction $x$.

To simplify the notation, we can rewrite eq.~(\ref{eq:ff}) as
\be\label{eq:PxNP}
\sigma_H(\cdot,Q) \simeq 
\sigma_Q(\cdot, Q,m) \otimes D^{np}_{Q\to H}(\cdot, \{params\}) \, ,
\ee
and this form allows one to read $\sigma_H(\cdot)$, $\sigma_Q(\cdot)$, and $D^{np}_{Q\to H}(\cdot)$ either as differential distributions in $x$ (i.e.\ considering $\cdot=x$, and $\sigma$ actually being $d\sigma/dx$), as above, or as Mellin
moments\footnote{We define the Mellin moments of a distribution $D(x)$ as
\be
D(N) \equiv \int_0^1 dx\, x^{N-1} D(x) \, .
\ee
}
 of the original distributions, with $\cdot = N$ and the convolution operation $\otimes$ becoming simply a product in moments space.

In this expression we have explicitly shown that the perturbative cross section for production of a heavy-quark $Q$ depends on its mass, and that the non-perturbative component  $D^{np}_{Q\to H}(\cdot, \{params\})$ can depend on a set of parameters that will be made explicit later. Of course, neither $\sigma_Q$ nor $D^{np}_{Q\to H}$ are physical observables, and the parameters of the latter will in particular depend on the details of the definition of the former.

\subsection{Fixed order}

In this paper we are concerned with evaluating $\sigma_Q$ up to next-to-next-to-leading order (NNLO) accuracy in QCD, and complementing the fixed-order calculation with up-to next-to-next-to-leading-logarithmic (NNLL) accuracy resummation of quasi-collinear and of soft emissions. 

To do so, we first introduce the approximation, pioneered by Mele and Nason~\cite{Mele:1990yq,Mele:1990cw}, of factorising the perturbative contribution into a process-specific short-distance cross section $\hat\sigma_i$ and a universal decay function\footnote{The nomenclature `decay function' was originally introduced in~\cite{Collins:1981uk,Collins:1981uw} to denote the fragmentation of a parton $i$ into a hadron $H$, time-like counterpart to the `parton distribution functions' (PDFs). While the use of the `PDFs' nomenclature has remained current, the one of `decay functions' has mostly run out of favour, `fragmentation functions' being often preferred. In this paper we will refrain from using the term `fragmentation function' for an unphysical object, preferring to reserve it for the observable-level cross section defined in eq.~(\ref{eq:ff}), and we will rather use the term `initial conditions' or `evolved initial conditions' or `perturbative fragmentation functions' - all found in the literature - for the decay functions.}
$D_{i\to Q}$, 
\be
\label{eq:ffsum}
\sigma_Q(\cdot,Q,m) = \sum_i \hat\sigma_i(\cdot,Q,\muf) \otimes D_{i\to Q}(\cdot,\muf,m) + {\cal
O}\left(\left(\frac{m}{Q}\right)^p\right) \, ,
\ee
with $p \geq 1$. In this expression $i$ denotes a partonic channel (quarks and gluon, or singlet, gluon and non-singlet, depending on the choice of the basis) and $\muf$ a factorisation scale\footnote{\label{foot:residual}Here and in the following we list in the arguments of the functions only the `real' scales on which they explicitly depend, while we avoid writing  also the `residual' scales, on which a quantity can still depend beyond a given perturbative order.}, taken to be of the order of $Q \gg m$. Both the $\hat\sigma_i$ and the $D_{i\to Q}$ are calculable in perturbation theory (the latter as long as the heavy quark mass is sufficiently larger than $\Lambda_{QCD}$), and they are known up to NNLO accuracy~\cite{Rijken:1996vr,Mitov:2006wy,Blumlein:2006rr,Melnikov:2004bm,Mitov:2004du}. Note that the choice of a factorisation scheme is implicit in these formulas: neither the $\hat\sigma_i$ nor the $D_{i\to Q}$ are physical objects, and depend on the scheme. In this paper we will use the \MSbar scheme, and leave to future work the study of alternative factorisation schemes.

We shall denote the fixed order accuracy as LO, NLO or NNLO for a perturbative expansion up to order $\as^0$, $\as^1$ or $\as^2$ respectively. Higher order calculations also introduce a running strong coupling $\as(\mur)$ and corresponding renormalisation scale $\mur$, and a beyond-accuracy dependency on the number of active flavours $n_f$ used in the renormalisation procedure.

\subsection{Collinear resummation}

In order to avoid the appearance of large unresummed logarithms $\as^n \log^k(\muf/m)$ of quasi-collinear origin, Dokshitzer-Gribov-Lipatov-Altarelli-Parisi (DGLAP) evolution equations can be used to evolve the perturbatively calculated decay functions $D_{i\to Q}$  from an initial scale $\muzf \simeq m$ up to a final scale $\muf \simeq Q$. We will use the DGLAP evolution implemented in MELA~\cite{Bertone:2015cwa} to evaluate
\be
D_{i\to Q}(\cdot,\muf,m) = \sum_j E_{ij}(\cdot,\muf,\muzf) \otimes D_{j\to Q}(\cdot,\muzf,m)\,,
\ee
where the $E_{ij}(\muf,\muzf)$ represent the evolution factors stemming
from the solution of the DGLAP evolution equations, leading to the resummation of collinear logs to leading~\cite{Altarelli:1977zs} ($\as^n\log^n(\muf/\muzf)$), next-to-leading~\cite{Curci:1980uw,Furmanski:1980cm} ($\as^n\log^{n-1}(\muf/\muzf)$), or next-to-next-to-leading \cite{Moch:2004pa, Vogt:2004mw} ($\as^n\log^{n-2}(\muf/\muzf)$) logarithmic accuracy. 

The use of the decay functions $D_{i\to Q}(\muzf,m)$, perturbatively calculated at ${\cal O}(\as)$ (next-to-leading order)~\cite{Mele:1990yq,Mele:1990cw} and ${\cal O}(\as^2)$ (next-to-next-to-leading order)~\cite{Melnikov:2004bm,Mitov:2004du} at $\muzf \simeq m$, as inputs for the DGLAP evolution justifies the name of `initial conditions' that is often employed  to describe them and that we will use from here on.

\subsection{Soft resummation}

Both the coefficient functions and the initial condition $D_{Q\to Q}$ exhibit large-logarithmic behaviour of soft origin in the $x\to 1$ region (or, equivalently, large-$N$ in moment space, $\as^n \log^k N$ with $k \leq 2n$). Such logarithms can be resummed to all orders, and results are known up to NNLL accuracy for the initial conditions~\cite{Cacciari:2001cw,Aglietti:2006yf,Maltoni:2022bpy,Czakon:2022pyz} and up to N3LL for the coefficient function~\cite{Cacciari:2001cw,Blumlein:2006pj,Moch:2009my}. Explicit expressions are given in Appendix~\ref{app:pscf} and \ref{app:psic}.

The resummed expressions exhibit a Landau pole related to the onset of
non-perturbative phenomena. In order to push the prediction above this point (i.e.\ to large $N$ or, equivalently, until $x=1$), such a pole must be regularised. Various options are available, they are detailed in Appendix \ref{app:landau}, and we will study the quantitative impact that they have on the final result. 

We denote a soft-resummed and regularised initial condition as
\be
D_{i\to Q}^{res, reg} \equiv \left(1 + \frac{\as}{2\pi} \bar d^{(1)} + 
                             \left(\frac{\as}{2\pi}\right)^2\bar d^{(2)}\right)\, D_{i\to Q}^{sud, reg}
\ee
where `$res$' denotes soft-resummation up to a given logarithmic accuracy, and `$reg$' the regularisation chosen for the Landau pole. $D_{i\to Q}^{res, reg}$ is defined in more detail in eq.~(\ref{eq:inires}), with the constants  $\bar d^{(1)}$ and $\bar d^{(2)}$ given in eqs.~(\ref{eq:iniconst1}) and (\ref{eq:iniconst2}). An equivalent expression for the coefficient function is defined in eqs.~(\ref{eq:cfres}), (\ref{eq:cfconst1}) and (\ref{eq:cfconst2}).

Finally, we $match$ the resummed expression to the fixed order ($fo$) one, so as to avoid double counting, and denote it as
\be
D_{i\to Q}^{fo+res,match,reg}\, .
\ee
We can perform either a simple additive matching, {\it match} = add,
\be
D_{i\to Q}^{fo+res,\mathrm{add},reg} = D_{i\to Q}^{fo} + D_{i\to Q}^{res, reg} - [D^{res(,reg)}_{i\to Q}]_{\as^p} \, ,
\ee
or a so-called log-R one~\cite{Catani:1992ua}, {\it match} = logR\footnote{The log-R matching is similar in spirit, but not identical, to the multiplicative matching employed in ref.~\cite{Czakon:2022pyz}. The term $[\log D_{i\to Q}^{fo}]_{\as^p}$ is equal to $(\as/2\pi) d^{(1)}$ at NLO and to $(\as/2\pi) d^{(1)} + (\as/2\pi)^2(-\frac{1}{2}{d^{(1)}}^2 + d^{(2)})$ at NNLO, where $d^{(1)}$ and $d^{(2)}$ are the perturbative coefficients in eq.~(\ref{eq:dini}). An equivalent expansion holds for the coefficient function, with the coefficients given in eq.~(\ref{CqCg})}
\be
\log D_{i\to Q}^{fo+res,\mathrm{logR},reg} = [\log D_{i\to Q}^{fo}]_{\as^p} + \log D_{i\to Q}^{sud, reg} - [\log D^{sud(,reg)}_{i\to Q}]_{\as^p} \, .
\ee
In these expressions, the notation $[...]_{\as^p}$ denotes expansion and truncation up to order $\as^p$, where $p=1$ for NLO accuracy and $p=2$ for NNLO accuracy. The `{\it (,reg)}' label in the expanded terms $[...]_{\as^p}$ is a reminder that whatever regularisation procedure may have been employed in the resummed expressions, it can also be applied to the expanded version, even if it is not necessary because the latter does not exhibit a Landau pole. The choice of using or not the regularisation in both terms amounts to a freedom contributing to an uncertainty stemming from this procedure.

An equivalent notation is used for the coefficient functions.

\subsection{Perturbative accuracy and scheme choices}

The numerical value of the overall prediction for a heavy quark fragmentation function will depend  on a number of choices, some discrete and some continuous ones:
\begin{enumerate}
\item
the accuracy of the fixed order calculation used in the coefficient functions and in the initial conditions. We denote it as `{\it fo}' and it can take the values `LO', `NLO' and `NNLO'
\item
\begin{enumerate}
\item
the logarithmic accuracy of the collinear resummation. We denote it as `$res_{coll}$' and it can take the values `LL', `NLL' and `NNLL'
\item
the logarithmic accuracy of the soft resummation. We denote it as `$res_{soft}$' and it can take the values `LL', `NLL' and `NNLL'
\end{enumerate}
Unless otherwise noted, we usually synchronise the soft and collinear resummation accuracies, $res_{coll} = res_{soft}$, and collectively denote them through the value of `{\it res}'.
\item
the scheme used for the matching of the fixed order and of the soft-resummed calculation. We denote it as `{\it match}' and can take the values `add' (for additive) and `logR' (for log-R matching)
\item
the choice of regularisation of the Landau pole. We denote it as `{\it reg}' and it can take the values `noreg' (for no regularisation), `CNO' (with an additional value for a parameter $f$)~\cite{Cacciari:2005uk}, `CGMP'~\cite{Czakon:2022pyz}. We will also briefly consider a modified CNO, to be denoted `CNOmod'. Appendix~\ref{app:landau} details the definitions of these regularisations.
\item
the values of the renormalisation and factorisation scale around $m$
and around $Q$, respectively $\muzr$, $\muzf$ and $\mur$, $\muf$, can
also affect the numerical value of the result.
\end{enumerate}
We do not include in this list, for instance, the choice of the active flavour number scheme,  and various options in the solution of the DGLAP equations as offered by MELA, for which  we make a single choice, to be detailed in Appendix \ref{app:melasetup}.

A final result for $\sigma_Q$ can hence be labeled as\footnote{We only exceptionally list here also the scales on which $\sigma_Q$ has only residual dependence at higher perturbative accuracy: $\mur$, $\muf$, $\muzr$ and $\muzf$. As previously explained in footnote~\ref{foot:residual}, we will not usually list them.}
\be
\sigma_Q^{fo+res,match,reg}(\cdot, Q,\mur, \muf, \muzr, \muzf,m) \, .
\ee

Our default (``best'') choice, to be justified below,  will be
\be
\sigma_Q^{\mathrm{NNLO+NNLL,logR,CNO(f=1.25)}}(\cdot, Q,\mur = \muf = Q, \muzr = \muzf = m,m)\, .
\ee

\section{Numerical Results}
\label{sec:numres}

We produce numerical results setting the pole masses of the heavy quarks to $m_c = 1.5$~GeV, $m_b = 4.75$~GeV and $m_t = 172.5$~GeV. The value of the strong coupling is set at a scale $Q=91.2$~GeV as $\as(Q=91.2~\text{GeV}) = 0.118$, and run from this initial condition to the scale where it is needed using an NNLO-accurate evolution equation in the \MSbar scheme with a variable number of active flavours. 

We evolve numerically all partons and the strong coupling using the package MELA~\cite{Bertone:2015cwa}. Details are given in Appendix~\ref{app:melasetup}.

We set the thresholds for changing the number of active flavours in a variable flavour number evolution scheme (VFNS) at the same values as the quark masses, i.e.\ we use 3 flavours below $m_c$, 4 flavours between $m_c$ and $m_b$, 5 flavours between $m_b$ and $m_t$, and 6 flavours above the top mass.\footnote{Since the initial conditions are always used in the $n_l = n_f-1$ scheme, when they are evaluated with an $\as^{(n_f)}$ value -- which can happen when one varies the renormalisation scale $\muzr$ slightly above $m$ -- a compensating factor for the value of $\as$ and the number of active flavours is introduced. The numerical impact of this procedure is anyway minimal.}

No other physical numerical parameter enters the perturbative predictions, other than the value of the centre-of-mass energy $Q$.

\subsection{Initial conditions}
\label{sec:ini}

Before considering the full, physical fragmentation functions, we study the initial conditions which, because of the low scales~$\simeq m$ at which they are evaluated (this is the case for the charm and the bottom quark), are the ingredient which is the most sensitive to the details of the perturbative setup.

We start by considering the bottom quark case, for easier comparison with the existing recent literature~\cite{Maltoni:2022bpy,Czakon:2022pyz}. In particular, we study in some detail the $D_{b\to b}(N,\muzf,m)$ component.

\begin{figure}[p]
\includegraphics[width=\textwidth]{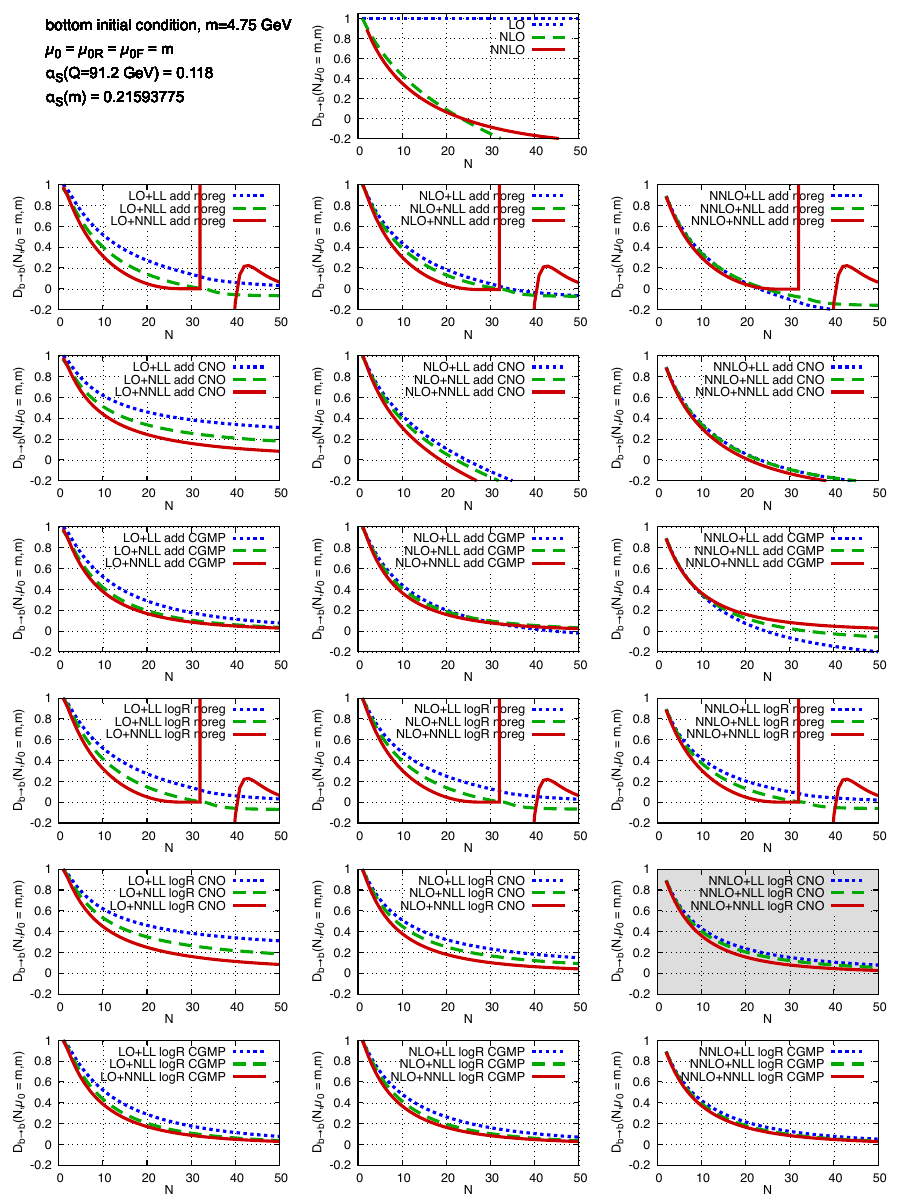}
\caption{\label{fig:bottom-ini}The initial condition for the bottom-to-bottom fragmentation function, without any evolution. `CNO' is always used here with the parameter $f=1.25$. The plot with a grey background contains the curve that will be used as a reference, i.e.\ `NNLO+NNLL logR CNO'.}
\end{figure}

\begin{figure}[p]
\includegraphics[width=\textwidth]{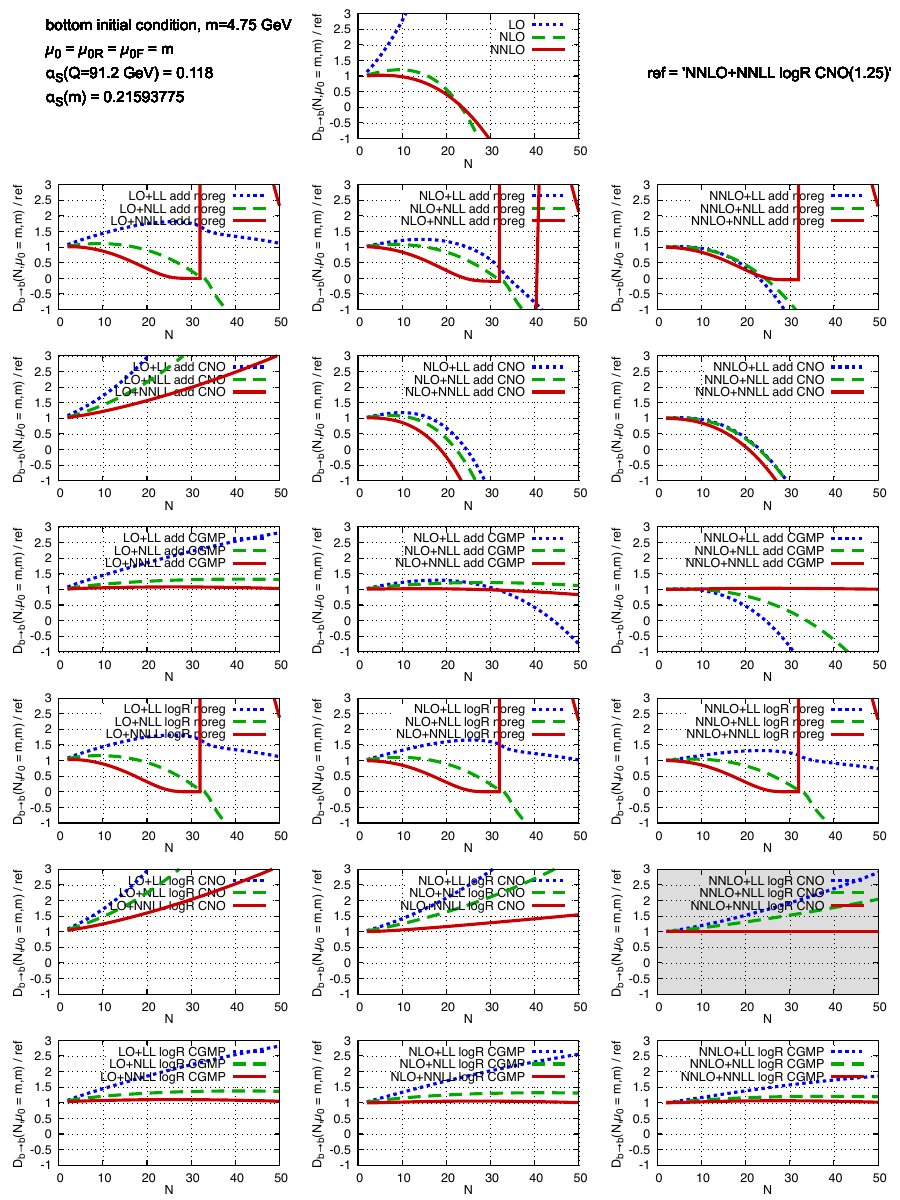}
\caption{\label{fig:bottom-ini-ratios}The initial condition for the bottom-to-bottom fragmentation function, without any evolution, normalised to the curve `NNLO+NNLL logR CNO' (the red curve in the plot with a grey background in figure~\protect\ref{fig:bottom-ini}). `CNO' is always used here with the parameter $f=1.25$. }
\end{figure}

Figure~\ref{fig:bottom-ini} displays, as a function of the Mellin moment index $N$, the moments of the function $D_{b\to b}(N,\mu_0=m,m)$ for the full range of perturbative orders, logarithmic accuracies of the soft resummation, matching schemes and regularisations of the Landau pole considered in this paper.\footnote{The discontinuities or irregularities in some of the curves denote the presence of the Landau pole for a specific value of $N=N^L_0$ in the resummed expressions, with the initial condition becoming infinite at that point. As expected from the expression $N^L_0 = \exp(1/(2b_0\as(m))$ (see Appendix~\ref{app:landau}), this value is $N^L_0 \simeq 30 $ for bottom quarks. Beyond $N^L_0$ the function becomes complex-valued, and what is plotted is its real part.}

Figure~\ref{fig:bottom-ini-ratios} displays, for better readability, the ratio of all the curves in figure~\ref{fig:bottom-ini} to the curve `NNLO+NNLL logR CNO', i.e.\ the red curve in the plot with a grey background in figure~\ref{fig:bottom-ini}.

The main take-away from this survey is that no clear and consistent hierarchy of perturbative orders (i.e.\ LO $>$ NLO $>$ NNLO) can be observed, nor a systematic convergence can be seen as the perturbative accuracy increases, either in the fixed order or in the resummed cases, the one exception seemingly being the CGMP regularisation of the Landau pole, with a log-R-type matching of the fixed-order and resummed results. This also holds true for the charm and the top quark cases (see figures~\ref{fig:charm-ini-ratios} and \ref{fig:top-ini-ratios} in Appendix~\ref{app:ini}). 

While it is tempting to conclude, on the basis of this observation, that CGMP is an ideal choice for the regularisation\footnote{While it is effective, one could argue that the CGMP regularisation of the Landau pole, based on a perturbative truncation of the exponent of the resummation factor that preserves the desired logarithmic accuracy, is no more strongly justified than other choices. In fact, it employs a perturbative recipe to provide a regularisation of a non-perturbative feature. Ref.~\cite{Czakon:2022pyz} also shows that -- contrary to the CNO regularisation introduced in~\cite{Cacciari:2005uk} -- it does not prevent the appearance of non-perturbative power corrections larger than expected.}, we will see that it makes it impossible to describe charmed meson data with the same simple non-perturbative functional forms that instead work well for the bottom meson data. This lack of universality will lead us to prefer using the CNO regularisation in this work.
It remains of course worth exploring if it is possible to combine the CGMP regularisation, and the better perturbative convergence it seems to lead to, with an effective non-perturbative parametrisation allowing for good fits to all the data.

\begin{figure}[t]
\includegraphics[width=\textwidth]{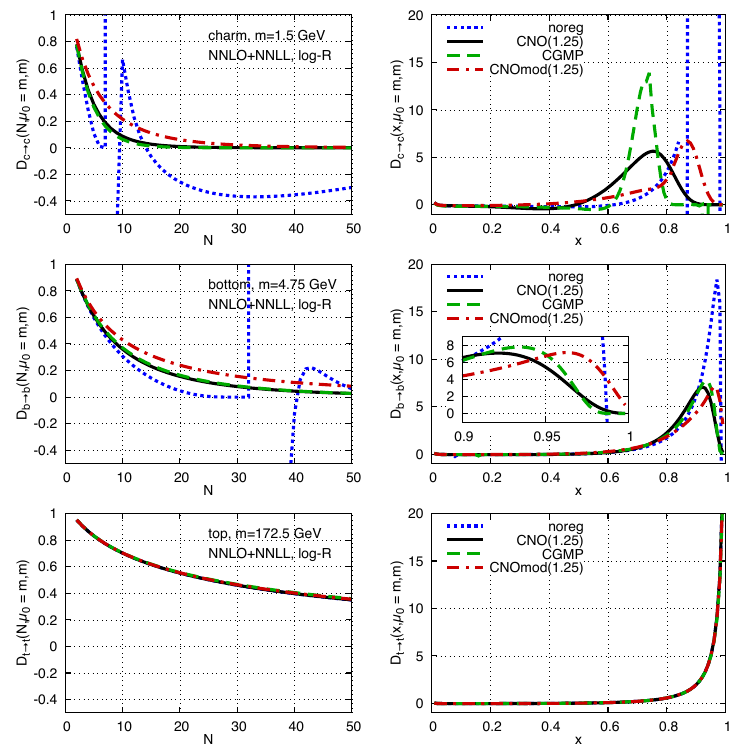}
\caption{\label{fig:x_and_CNOmod} $N$-space (left column) and $x$-space (right column) representation of the perturbative initial conditions for the three heavy quarks, evaluated at NNLO+NNLL accuracy with log-R matching, with different kinds of Landau pole regularisations. In all cases $\muzf = \muzr = m$, and $\as(91.2~\mathrm{GeV}) = 0.118$. CNO and CNOmod are always implemented with $f=1.25$. }
\end{figure}

In order to better appreciate the impact of different non-perturbative regularisations of the Landau pole on the otherwise perturbative initial conditions, we show in figure~\ref{fig:x_and_CNOmod} the NNLO+NNLL curve with log-R matching with different regularisations and for all three heavy quarks, both in $N$ and in $x$ space. We see that the resulting curves can be very different, especially in the large-$x$ region. The non-perturbative contributions that will have to be added in order to fit the experimental data, where available, will consequently also be very different. In some cases, the shape of the perturbative fragmentation function can make it impossible to fit the experimental data with simple non-perturbative forms: we will see that this is notably the case for charmed hadron data and CGMP regularisation, which leads to a flat and null initial condition in a large region $x \gtrsim 0.8$ (this characteristic remains after evolution and convolution with the coefficient function).

In figure~\ref{fig:x_and_CNOmod} we have also introduced a new regularisation for the Landau pole, dubbed `CNOmod', which is meant to be CNO but implemented as it is presumably done in ref.~\cite{Czakon:2022pyz}, i.e.\ rescaling only the resummation term but not the subtraction one (which is devoid of a Landau pole, since it is expanded at a fixed order in $\as$) in the matching (see appendix~\ref{app:landau}). We can see that the feature of the initial condition with CNO not vanishing for $x\to 1$, pointed out in ref.~\cite{Czakon:2022pyz} as being problematic (cfr.~figure 2 there and the text below) may in fact stem from the specific implementation of CNO that that work adopts.\footnote{Ref.~\cite{Czakon:2022pyz} uses $f=1.25$, like \cite{Cacciari:2005uk}. If $f$ is chosen closer to one, the vanishing of the initial condition at $x=1$ improves, but at the expense of a less effective screening of the Landau pole.} When CNO is implemented as in~\cite{Cacciari:2005uk}, rescaling the resummed term but also its expansion, the initial condition appears to vanish quickly when $x\to 1$ (see also Appendix~\ref{app:landau}).

\begin{figure}[t]
  \includegraphics[width=\textwidth]{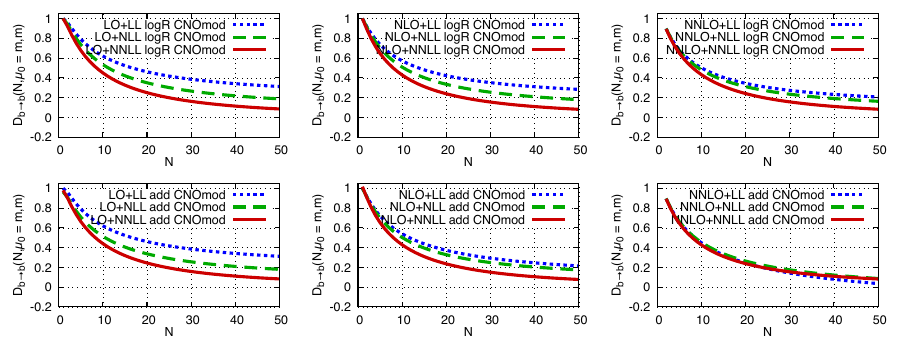}
  \caption{\label{fig:bottom-ini-CNOmod}The initial condition for the bottom-to-bottom fragmentation function, without any evolution. `CNOmod' is always used here with the parameter $f=1.25$. As in figure~\ref{fig:bottom-ini}, $m=4.75$~GeV, $\mu_0=\muzr=\muzf=m$, $\as(91.2~{\mathrm{GeV}})=0.118$, $\as(m)=0.21593775$.}
\end{figure}

It is also worth noting how, in figures \ref{fig:bottom-ini} and \ref{fig:bottom-ini-ratios}, the plots employing additive matching and CNO regularisation of the Landau pole with NLO and NNLO fixed order appear to be outliers to some extent, with the moments becoming very rapidly small and eventually negative at large $N$. This behaviour can eventually be traced back to the feature of the CNO regularisation of rescaling both the resummation term and the subtraction one: this leaves the (unrescaled) logarithms present in the fixed order partially unsubtracted, leading to the observed patterns. Employing the CNOmod regularisation instead of the CNO one, i.e. only rescaling the resummation term, prevents this behaviour, as shown in figure~\ref{fig:bottom-ini-CNOmod}. In this work we have nevertheless chosen CNO as our default regularisation of the Landau pole, in part because it better preserves the exact fixed order calculation at small $N$, and in part because of easier comparison with ref.~\cite{Cacciari:2005uk}. When employing a log-R rather than an additive matching, which will also be our default, the difference is minimal anyway.

\subsection[Full $e^+e^-$ fragmentation function]{Full \boldmath{$e^+e^-$} fragmentation function}

The full fragmentation function in $e^+e^-$ collisions is obtained convoluting the  coefficient functions with the evolved initial conditions, summing over all parton channels, as shown in eq.~(\ref{eq:ffsum}). We present here numerical results for fragmentation to bottom quarks, $\sigma_b$, and we study theoretical uncertainties stemming from missing higher perturbative orders by varying the renormalisation and the factorisation scales, both around the initial scale $\mu_0$, set equal to the heavy quark mass, and around the final scale $Q$, the centre-of-mass energy of the collision. More specifically, we will consider
\be
Q/2 \le \mur,\muf \le 2Q, 
\quad\mathrm{with}\quad 1/2 \le \mur/\muf \le 2
\ee
for the coefficient functions, and
\be
m/2 \le \muzr \le 2m\quad\mathrm{and}\quad m \le \muzf \le 2m, 
\quad\mathrm{with}\quad 1/2 \le \muzr/\muzf \le 2
\ee
for the initial conditions, resulting in the customary 7-point rule in the first case, and in a 5-point rule in the second.\footnote{
The choice of limiting the variation of $\muzf$ to scales larger than the heavy quark mass $m$ is due to a technical limitation of the MELA package, which does not allow for evolution in a Variable Flavour Number Scheme with $n$ flavours from a starting scale below the mass of the $n^{th}$ flavour. While this could in principle be fixed, the remaining variations already give a sufficiently realistic picture of the uncertainty, and in most practical cases the initial scales will be fixed at $m$ anyway. Lifting this limitation was therefore considered not to be a priority.}

\begin{figure}[t]
\includegraphics[width=\textwidth]{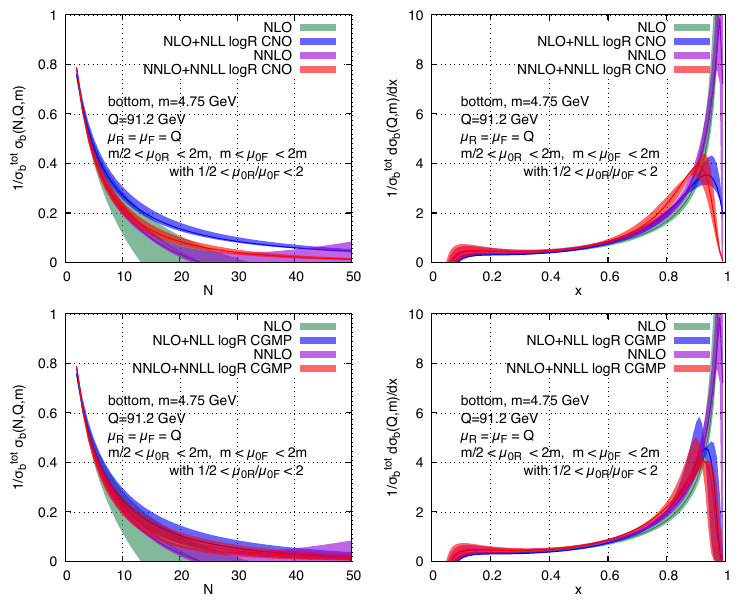}
\caption{\label{fig:ff-bottom-ini} Bottom fragmentation function. Uncertainty bands due to variation of the $\muzr$ and $\muzf$ scales around $m$. Left plots in $N$-space, right plots in $x$-space. Upper plots with CNO regularisation ($f=1.25$), lower plots with CGMP regularisation.}
\end{figure}

Figure~\ref{fig:ff-bottom-ini} shows the uncertainty bands, in $N$- and in $x$-space, due to $\muzr$ and $\muzf$ variation in the region detailed above, for four different perturbative approximations, NLO, NNLO, NLO+NLL resummation (collinear and soft) and NNLO+NNLL resummation (collinear and soft), the latter two with log-R matching and CNO or CGMP Landau pole regularisation. 
In the $N$-space (left) plots, one can see some degree of improvement (i.e.\ narrower bands) as the accuracy of the perturbative series increases. However, on one hand the NNLO+NNLL band (red) is not significantly narrower than the NLO+NLL (blue) one, and on the other hand the two bands are hardly or not at all overlapping (especially in the CNO regularisation case), indicating poor convergence of the series. This poor convergence, and the significant differences between the CNO and the CGMP cases, are consistent with the observations made in section~\ref{sec:ini}.
One last consideration that we make is that the shape and the position of the bands in $x$-space do not tell us much about convergence. Moreover, they can drastically depend on the chosen regularisation for the Landau pole, again as observed in section~\ref{sec:ini}.

\begin{figure}[t]
\includegraphics[width=\textwidth]{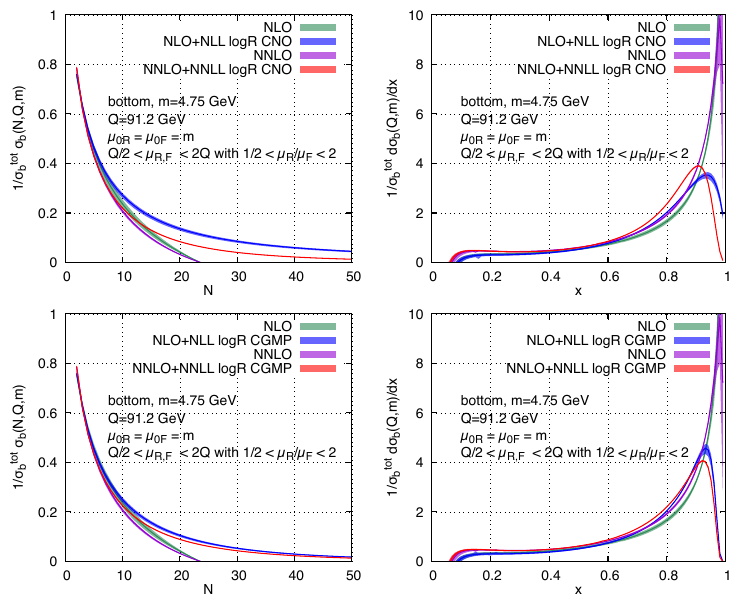}
\caption{\label{fig:ff-bottom} Bottom fragmentation function. Uncertainty bands due to variation of the $\mur$ and $\muf$ scales around $Q$. Left plots in $N$-space, right plots in $x$-space. Upper plots with CNO regularisation ($f=1.25$), lower plots with CGMP regularisation.}
\end{figure}

In figure~\ref{fig:ff-bottom} we repeat the same exercise, but varying this time the scales $\mur$ and $\muf$ around $Q$. The bands are much narrower, due to the value of $\as$ being significantly smaller around the large scale $Q=91.2$~GeV than around the mass scale $m=4.75$~GeV. Because of this smaller value of $\as$, the expected perturbative hierarchy is better respected, with the bands becoming generally narrower as the perturbative precision increases. As was the case in figure~\ref{fig:ff-bottom-ini}, bands in $x$-space are instead not easily interpretable in terms of behaviour of the perturbative series.

Analogous figures for the charm and top quark cases can be found in Appendix~\ref{app:fullff}.

\section{Comparisons to Experimental Data}
\label{sec:fits}

In this section we study how the fragmentation functions that we have calculated, possibly complemented by a non-perturbative component, fare in describing $D$ and $B$ meson fragmentation data measured in $e^+e^-$ collisions.
We consider the following datasets: $B$-mesons production data at the $Z$ peak ($Q=91.2$~GeV), as measured by the ALEPH~\cite{ALEPH:2001pfo}, OPAL~\cite{OPAL:2002plk}, SLD~\cite{SLD:2002poq} and DELPHI~\cite{Barker:2002iuq} collaborations\footnote{While the ALEPH set refers specifically to B-mesons, the SLD, DELPHI and OPAL data are for all weakly decaying bottom-flavoured hadrons and include about 10\% of baryons.}; $D$-mesons production data at the $Z$ peak, as measured by the ALEPH~\cite{ALEPH:1999syy} collaboration; $D$-mesons production data near the $\Upsilon(4S)$ peak ($Q=10.6$~GeV), as measured by the CLEO~\cite{CLEO:2004enr} and BELLE~\cite{Belle:2005mtx} collaborations.\footnote{\label{foot:belle} The BELLE data from ref.~\cite{Belle:2005mtx} have been removed from HEPData in 2014 `at the request of the authors due to an unrecoverable errror in the measurement' (see \url{https://www.hepdata.net/record/ins686014}). However, since the corresponding paper does not seem to have been formally retracted, and the CLEO data appear compatible where comparable, we have decided to keep using these BELLE data in this paper, in part to facilitate comparisons with ref.~\cite{Cacciari:2005uk}.}
Electromagnetic initial-state radiation (ISR) effects are accounted for in the $D$-mesons production data near the $\Upsilon(4S)$ peak, according to the procedure detailed in section 3 of~\cite{Cacciari:2005uk}.

\begin{figure}[t]
  \includegraphics[width=0.49\textwidth]{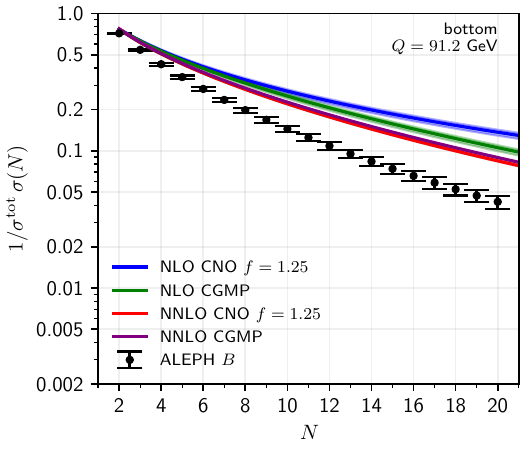}
  \includegraphics[width=0.49\textwidth]{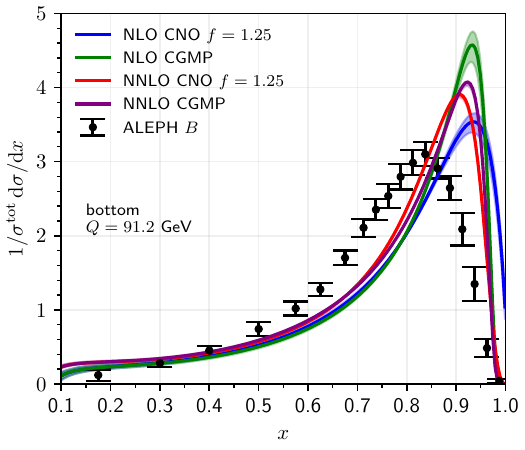}
  \caption{Perturbative predictions for bottom quark production at $Q=91.2$~GeV compared to ALEPH $B$-meson data. `NLO' stands for `NLO+NLL logR', while `NNLO' stands for `NNLO+NNLL logR'.}
  \label{fig:pertB}
\end{figure}

\begin{figure}[t]
  \includegraphics[width=0.49\textwidth]{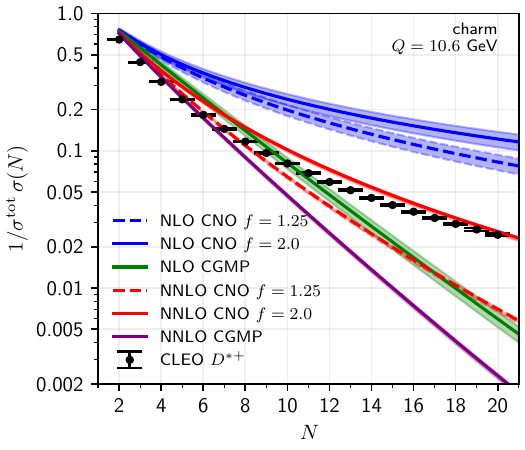}
  \includegraphics[width=0.49\textwidth]{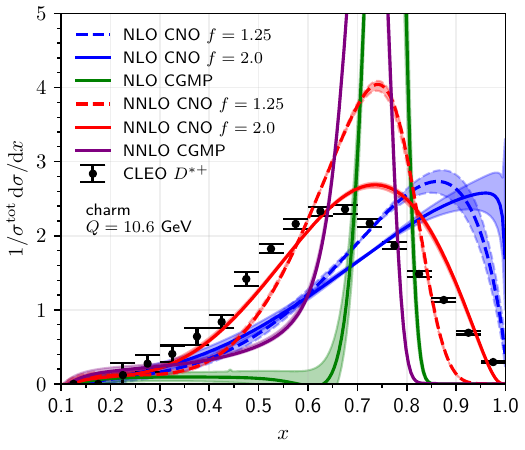}
  \caption{Perturbative predictions for charm quark production at $Q=10.6$~GeV compared to CLEO $D^{*+}$-meson data. `NLO' stands for `NLO+NLL logR', while `NNLO' stands for `NNLO+NNLL logR'.}
  \label{fig:pertC}
\end{figure}

We first show how perturbative curves for different perturbative orders (NLO+NLL and NNLO+NNLL) and different prescriptions for the regularisation of the Landau pole (CNO, CGMP) compare to representative data points for $B$-meson production in figure~\ref{fig:pertB} and $D$-meson production in figure~\ref{fig:pertC}.
We present curves both in Mellin and in direct space, respectively on the left and the right panels.

As already stressed in section~\ref{sec:numres}, for the bottom quark the behaviour of CNO and CGMP is similar at NNLO, and both results can be easily used as perturbative contributions to fit a non-perturbative component to the data.
For the charm quark instead, CNO and CGMP behave qualitatively and quantitatively differently, both at NLO and at NNLO. In direct space (figure~\ref{fig:pertC}, right panel), we see that both CNO with $f=1.25$ and CGMP feature a sharp peak around $x\sim$~0.7--0.8 and a steep drop around $x\sim$~0.8--0.9. In moment space (figure~\ref{fig:pertC}, left panel), such behaviour translates to the NNLO curves dipping below the experimental data points at $N\gtrsim$~6.
This fact immediately prevents one from fitting the data using the NNLO result and simple non-perturbative functional forms: their Mellin moments being normalised to values smaller than one (because these non-perturbative forms are meant to degrade the quark momentum when it fragments into a hadron), they can only lower the value of the theoretical prediction, but not raise it.
A simple practical solution to this problem is to increase the value of the free parameter of the CNO regularisation to $f=2$. Figure~\ref{fig:pertC} shows that, in this case, for the charm quark the NNLO+NNLL curves remains above the data in Mellin space up to larger values of N (left panel), suggesting that a simple non-perturbative form may be able to describe the data. The right panel in the same figure shows that, indeed, the $x$ distribution is closer to the shape of the data.

\subsection{Single-moment and single-parameter fits}
\label{sec:spf}

\begin{figure}[t]
  \centering
  \includegraphics[width=0.8\textwidth]{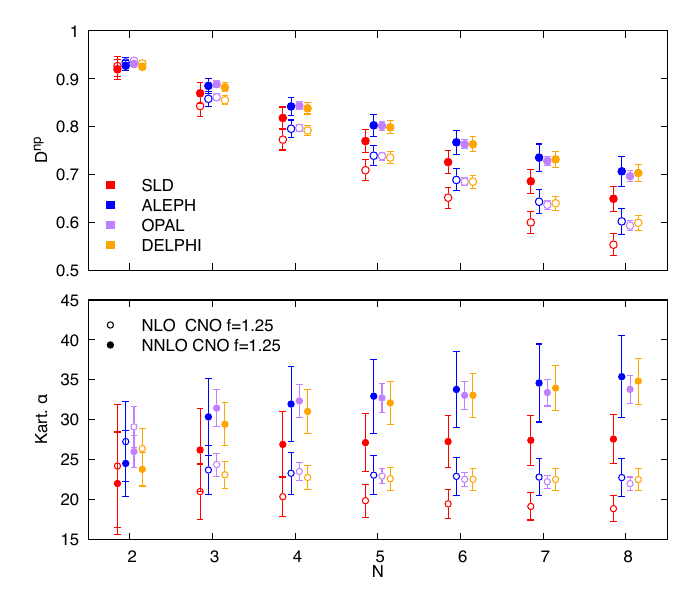}
  \caption{\label{fig:spfitsB} Single-point fits to $B$-meson data from experiments at the $Z$ peak, with the one-parameter non-perturbative expression of Kartvelishvili et al.\ in eq.~(\ref{eq:kart}). Upper pane: value of normalised non-perturbative moment, fitted to a single, specific moment. Lower pane: corresponding value of $\alpha$. NLO stands for 'NLO+NLL logR', while NNLO stands for 'NNLO+NNLL logR'.}
\end{figure}

\begin{figure}[t]
  \centering
  \includegraphics[width=0.8\textwidth]{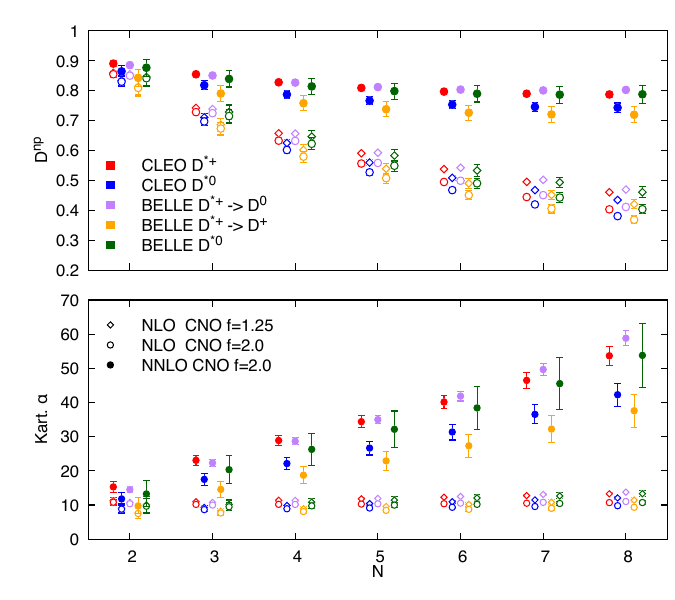}
  \caption{\label{fig:spfitsD} Single-point fits to $D$-mesons data from experiments at the $\Upsilon(4S)$ peak, with the one-parameter non-perturbative expression of Kartvelishvili et al.\ in eq.~(\ref{eq:kart}). Upper pane: value of normalised non-perturbative moment, fitted to a single, specific moment. Lower pane: corresponding value of $\alpha$. NLO stands for 'NLO+NLL logR', while NNLO stands for 'NNLO+NNLL logR'.}
\end{figure}

Here we try to describe single moments of experimental data using a very simple non-perturbative functional form, dependent on a single parameter, on top of our perturbative description. We use the form introduced by Kartvelishvili et al.~\cite{Kartvelishvili:1977pi}
\be
\label{eq:kart}
D^{np}_K(x) = (\alpha + 1)(\alpha + 2)x^\alpha(1-x) \, ,
\ee
whose Mellin transform is
\be\label{eq:KartN}
D^{np}_K(N) = \frac{(\alpha + 1)(\alpha + 2)}{(\alpha+N)(\alpha+N+1)} \, .
\ee
In the following we present our results for the values of the parameter $\alpha$ for $B$ and $D$-mesons data.
We determine the size of the non-perturbative component in Mellin space by dividing the moments of the experimental data by the perturbative result according to eq.~\eqref{eq:PxNP}.

The result for $B$ mesons is shown in the
upper panel of figure~\ref{fig:spfitsB}, for values of $N$ between 2 and 8. This range has been chosen because, if one were to use the results of these fits in hadronic collisions, where steeply falling transverse momentum distribution~$\sim 1/p_T^5$ would have to be convoluted, the moments at $N\simeq 5$ would provide the dominant contribution to the result.

In the lower panel of figure~\ref{fig:spfitsB}, for each moment $N$, we show the
value of $\alpha$ determined by inverting eq.~\eqref{eq:KartN}.
We notice that both at NLO+NLL and NNLO+NNLL the extracted values of $\alpha$ are fairly consistent at different values of $N$, showing that this simple non-perturbative form manages to describe fairly well the shape of the whole distribution.
We also observe that the larger value of $\alpha$ extracted at NNLO than at NLO is consistent with the comparison shown in figure~\ref{fig:pertB}, since the NNLO curve is closer to the data and needs therefore to be supplemented with a smaller non-perturbative component (i.e.\ a harder function, i.e.\ a larger $\alpha$).

We consider similar fits to $D$ meson data. The results are shown in figure~\ref{fig:spfitsD} where we consider experimental data for $D^{*+}$ and $D^{*0}$ production\footnote{Analogous qualitative conclusions to the ones reported in this section hold for $D^{+}$ and $D^{0}$ mesons. We do not detail such results here for brevity.}.
At NLO+NLL, the  data can be described quite well with a single value
of $\alpha$ within uncertainties, independently of the value of $f$ used. Instead, and  at variance also with the single-moment
$B$ meson results of figure~\ref{fig:spfitsB}, this is not the case at NNLO+NNLL, where
a large dependence of $\alpha$ on $N$ can be observed in the lower panel of figure~\ref{fig:spfitsD}.
We also observe that at NNLO level it is mandatory to choose $f=2.0$, as shown by the perturbative result/data comparison in figure~\ref{fig:pertC}.
This points to a stronger dependence of the non-perturbative component of the $D$ meson fragmentation on the details of the Landau pole regularisation applied to the NNLO+NNLL perturbative result.

Numerical results for $D^{np}$ and for the $\alpha$ parameter of eq.~(\ref{eq:kart}) for $B$ and $D$ mesons are collected in tables~\ref{tab:DNP} and \ref{tab:alpha}. At NLO+NLL level they are not identical to those presented in table 4 of~\cite{Cacciari:2005uk}, but we have checked that this is exclusively an effect of having used different perturbative parameters (essentially the value of $\as$). This is of course legitimate, the non-perturbative component being unphysical by itself and dependent on the setup used in the perturbative part.

\begin{table}
\begin{sideways}
\begin{minipage}{\textheight}
\small
\begin{tabular}{|l|c|c|c|c|c|c|c|}
\hline
$N$ & 2 & 3 & 4 & 5 & 6 & 7 & 8 \\
\hline
\hline
\multicolumn{8}{|c|}{$D^{np}(N)$} \\
\hline
\hline
\multicolumn{8}{|c|}{NLO+NLL logR CNO($f=1.25$)} \\
\hline
SLD $B$ &
$0.93 \pm 0.02$ &
$0.84 \pm 0.02$ &
$0.77 \pm 0.02$ &
$0.71 \pm 0.02$ &
$0.65 \pm 0.02$ &
$0.60 \pm 0.02$ &
$0.55 \pm 0.02$ \\
ALEPH $B$ &
$0.93 \pm 0.01$ &
$0.86 \pm 0.02$ &
$0.80 \pm 0.02$ &
$0.74 \pm 0.02$ &
$0.69 \pm 0.02$ &
$0.64 \pm 0.02$ &
$0.60 \pm 0.03$  \\ 
OPAL $B$ &
$0.938 \pm 0.005$ &
$0.86 \pm 0.01$ &
$0.80 \pm 0.01$ &
$0.74 \pm 0.01$ &
$0.68 \pm 0.01$ &
$0.64 \pm 0.01$ &
$0.59 \pm 0.01$ \\
DELPHI $B$ &
$0.93 \pm 0.01$ &
$0.86 \pm 0.01$ &
$0.79 \pm 0.01$ &
$0.74 \pm 0.01$ &
$0.68 \pm 0.01$ &
$0.64 \pm 0.01$ &
$0.60 \pm 0.02$ \\
\hline
CLEO $D^{*+}$ &
$0.86 \pm 0.01$ &
$0.74 \pm 0.01$ &
$0.66 \pm 0.01$ &
$0.59 \pm 0.01$ &
$0.54 \pm 0.01$ &
$0.50 \pm 0.01$ &
$0.46 \pm 0.01$ \\
CLEO $D^{*0}$ &
$0.83 \pm 0.02$ &
$0.71 \pm 0.01$ &
$0.63 \pm 0.01$ &
$0.56 \pm 0.01$ &
$0.51 \pm 0.01$ &
$0.47 \pm 0.01$ &
$0.43 \pm 0.01$ \\
BELLE $D^{*+} \to D^0$ &
$0.85 \pm 0.01$ &
$0.739 \pm 0.005$ &
$0.657 \pm 0.004$ &
$0.593 \pm 0.004$ &
$0.542 \pm 0.004$ &
$0.502 \pm 0.004$ &
$0.469 \pm 0.004$ \\
BELLE $D^{*+} \to D^+$ &
$0.81 \pm 0.03$ &
$0.69 \pm 0.02$ &
$0.60 \pm 0.02$ &
$0.54 \pm 0.02$ &
$0.49 \pm 0.02$ &
$0.45 \pm 0.02$ &
$0.42 \pm 0.02$ \\
BELLE $D^{*0}$ &
$0.85 \pm 0.03$ &
$0.73 \pm 0.02$ &
$0.65 \pm 0.02$ &
$0.58 \pm 0.02$ &
$0.53 \pm 0.02$ &
$0.49 \pm 0.02$ &
$0.46 \pm 0.02$ \\
\hline
\multicolumn{8}{|c|}{NLO+NLL logR CNO($f=2.0$)} \\
\hline
CLEO $D^{*+}$ &
$0.86 \pm 0.01$ &
$0.73 \pm 0.01$ &
$0.63 \pm 0.01$ &
$0.56 \pm 0.01$ &
$0.495 \pm 0.005$ &
$0.445 \pm 0.005$ &
$0.403 \pm 0.005$ \\
CLEO $D^{*0}$ &
$0.83 \pm 0.02$ &
$0.70 \pm 0.01$ &
$0.60 \pm 0.01$ &
$0.53 \pm 0.01$ &
$0.47 \pm 0.01$ &
$0.42 \pm 0.01$ &
$0.38 \pm 0.01$ \\
BELLE $D^{*+} \to D^0$ &
$0.85 \pm 0.01$ &
$0.725 \pm 0.005$ &
$0.632 \pm 0.004$ &
$0.558 \pm 0.004$ &
$0.499 \pm 0.003$ &
$0.451 \pm 0.003$ &
$0.411 \pm 0.003$ \\
BELLE $D^{*+} \to D^+$ &
$0.81 \pm 0.03$ &
$0.67 \pm 0.02$ &
$0.58 \pm 0.02$ &
$0.51 \pm 0.02$ &
$0.45 \pm 0.02$ &
$0.41 \pm 0.01$ &
$0.37 \pm 0.01$ \\
BELLE $D^{*0}$ &
$0.84 \pm 0.03$ &
$0.72 \pm 0.02$ &
$0.62 \pm 0.02$ &
$0.55 \pm 0.02$ &
$0.49 \pm 0.02$ &
$0.44 \pm 0.02$ &
$0.40 \pm 0.02$ \\
\hline
\hline
\multicolumn{8}{|c|}{NNLO+NNLL logR CNO($f=1.25$)} \\
\hline
SLD $B$ &
$0.92 \pm 0.02$ &
$0.87 \pm 0.02$ &
$0.82 \pm 0.02$ &
$0.77 \pm 0.02$ &
$0.73 \pm 0.02$ &
$0.69 \pm 0.03$ &
$0.65 \pm 0.03$ \\
ALEPH $B$ &
$0.93 \pm 0.01$ &
$0.89 \pm 0.02$ &
$0.84 \pm 0.02$ &
$0.80 \pm 0.02$ &
$0.77 \pm 0.03$ &
$0.74 \pm 0.03$ &
$0.71 \pm 0.03$ \\
OPAL $B$ &
$0.931 \pm 0.005$ &
$0.89 \pm 0.01$ &
$0.84 \pm 0.01$ &
$0.80 \pm 0.01$ &
$0.76 \pm 0.01$ &
$0.73 \pm 0.01$ &
$0.70 \pm 0.01$ \\
DELPHI $B$ &
$0.93 \pm 0.01$ &
$0.88 \pm 0.01$ &
$0.84 \pm 0.01$ &
$0.80 \pm 0.01$ &
$0.76 \pm 0.02$ &
$0.73 \pm 0.02$ &
$0.70 \pm 0.02$ \\
\hline
\multicolumn{8}{|c|}{NNLO+NNLL logR CNO($f=2.0$)} \\
\hline
CLEO $D^{*+}$ &
$0.89 \pm 0.01$ &
$0.86 \pm 0.01$ &
$0.83 \pm 0.01$ &
$0.81 \pm 0.01$ &
$0.80 \pm 0.01$ &
$0.79 \pm 0.01$ &
$0.79 \pm 0.01$ \\
CLEO $D^{*0}$ &
$0.86 \pm 0.02$ &
$0.82 \pm 0.01$ &
$0.79 \pm 0.01$ &
$0.77 \pm 0.01$ &
$0.75 \pm 0.01$ &
$0.75 \pm 0.01$ &
$0.74 \pm 0.02$ \\
BELLE $D^{*+} \to D^0$ &
$0.89 \pm 0.01$ &
$0.85 \pm 0.01$ &
$0.83 \pm 0.01$ &
$0.81 \pm 0.01$ &
$0.80 \pm 0.01$ &
$0.80 \pm 0.01$ &
$0.80 \pm 0.01$ \\
BELLE $D^{*+} \to D^+$ &
$0.84 \pm 0.03$ &
$0.79 \pm 0.03$ &
$0.76 \pm 0.02$ &
$0.74 \pm 0.02$ &
$0.73 \pm 0.02$ &
$0.72 \pm 0.03$ &
$0.72 \pm 0.03$ \\
BELLE $D^{*0}$ &
$0.88 \pm 0.03$ &
$0.84 \pm 0.03$ &
$0.81 \pm 0.03$ &
$0.80 \pm 0.03$ &
$0.79 \pm 0.03$ &
$0.79 \pm 0.03$ &
$0.79 \pm 0.03$ \\
\hline
\end{tabular}
\caption{\label{tab:DNP}
Values of the first eight moments of the non-perturbative fragmentation function $D^{np}(N)$, as extracted from the experimental data.}
\end{minipage}
\end{sideways}
\end{table}

\begin{table}
\begin{sideways}
\begin{minipage}{\textheight}
\small
\centering
\begin{tabular}{|l|c|c|c|c|c|c|c|}
\hline
$N$ & 2 & 3 & 4 & 5 & 6 & 7 & 8 \\
\hline
\hline
\multicolumn{8}{|c|}{$\alpha$} \\
\hline
\hline
\multicolumn{8}{|c|}{NLO+NLL logR CNO($f=1.25$)} \\
\hline
SLD $B$ &
$24.18 \pm 7.69$ &
$20.97 \pm 3.46$ &
$20.33 \pm 2.53$ &
$19.82 \pm 2.09$ &
$19.43 \pm 1.84$ &
$19.10 \pm 1.70$ &
$18.83 \pm 1.61$ \\
ALEPH $B$ &
$27.24 \pm 5.05$ &
$23.67 \pm 3.06$ &
$23.27 \pm 2.63$ &
$23.03 \pm 2.45$ &
$22.88 \pm 2.38$ &
$22.79 \pm 2.37$ &
$22.73 \pm 2.39$ \\
OPAL $B$ &
$29.09 \pm 2.54$ &
$24.36 \pm 1.44$ &
$23.49 \pm 1.16$ &
$22.91 \pm 0.99$ &
$22.51 \pm 0.88$ &
$22.20 \pm 0.83$ &
$21.97 \pm 0.82$ \\
DELPHI $B$ &
$26.37 \pm 2.54$ &
$23.08 \pm 1.72$ &
$22.74 \pm 1.55$ &
$22.58 \pm 1.46$ &
$22.51 \pm 1.41$ &
$22.48 \pm 1.38$ &
$22.47 \pm 1.36$ \\
\hline
CLEO $D^{*+}$ &
$11.25 \pm 0.95$ &
$10.97 \pm 0.42$ &
$11.38 \pm 0.31$ &
$11.81 \pm 0.27$ &
$12.27 \pm 0.26$ &
$12.76 \pm 0.25$ &
$13.28 \pm 0.26$ \\
CLEO $D^{*0}$ &
$9.09 \pm 1.29$ &
$9.26 \pm 0.62$ &
$9.85 \pm 0.47$ &
$10.40 \pm 0.42$ &
$10.94 \pm 0.41$ &
$11.49 \pm 0.41$ &
$12.05 \pm 0.42$ \\
BELLE $D^{*+} \to D^0$ &
$10.77 \pm 0.53$ &
$10.76 \pm 0.27$ &
$11.34 \pm 0.21$ &
$11.91 \pm 0.19$ &
$12.49 \pm 0.18$ &
$13.10 \pm 0.18$ &
$13.74 \pm 0.18$ \\
BELLE $D^{*+} \to D^+$ &
$7.74 \pm 1.61$ &
$8.19 \pm 0.90$ &
$8.90 \pm 0.74$ &
$9.56 \pm 0.68$ &
$10.19 \pm 0.66$ &
$10.81 \pm 0.66$ &
$11.44 \pm 0.68$ \\
BELLE $D^{*0}$ &
$10.01 \pm 2.34$ &
$10.18 \pm 1.28$ &
$10.83 \pm 1.03$ &
$11.44 \pm 0.92$ &
$12.04 \pm 0.88$ &
$12.66 \pm 0.87$ &
$13.30 \pm 0.88$ \\
\hline
\multicolumn{8}{|c|}{NLO+NLL logR CNO($f=2.0$)} \\
\hline
CLEO $D^{*+}$ &
$10.82 \pm 0.89$ &
$10.17 \pm 0.37$ &
$10.20 \pm 0.26$ &
$10.27 \pm 0.22$ &
$10.37 \pm 0.20$ &
$10.52 \pm 0.19$ &
$10.70 \pm 0.19$ \\
CLEO $D^{*0}$ &
$8.79 \pm 1.22$ &
$8.65 \pm 0.56$ &
$8.90 \pm 0.41$ &
$9.12 \pm 0.35$ &
$9.34 \pm 0.32$ &
$9.56 \pm 0.31$ &
$9.79 \pm 0.31$ \\
BELLE $D^{*+} \to D^0$ &
$10.37 \pm 0.50$ &
$9.99 \pm 0.24$ &
$10.16 \pm 0.18$ &
$10.34 \pm 0.15$ &
$10.55 \pm 0.14$ &
$10.77 \pm 0.13$ &
$11.03 \pm 0.13$ \\
BELLE $D^{*+} \to D^+$ &
$7.50 \pm 1.53$ &
$7.68 \pm 0.82$ &
$8.09 \pm 0.64$ &
$8.43 \pm 0.56$ &
$8.74 \pm 0.53$ &
$9.04 \pm 0.51$ &
$9.33 \pm 0.51$ \\
BELLE $D^{*0}$ &
$9.65 \pm 2.20$ &
$9.48 \pm 1.14$ &
$9.74 \pm 0.87$ &
$9.97 \pm 0.75$ &
$10.19 \pm 0.68$ &
$10.45 \pm 0.65$ &
$10.71 \pm 0.64$ \\
\hline
\hline
\multicolumn{8}{|c|}{NNLO+NNLL logR CNO($f=1.25$)} \\
\hline
SLD $B$ &
$21.98 \pm 6.46$ &
$26.19 \pm 5.18$ &
$26.88 \pm 4.16$ &
$27.11 \pm 3.61$ &
$27.24 \pm 3.30$ &
$27.40 \pm 3.13$ &
$27.55 \pm 3.04$ \\
ALEPH $B$ &
$24.50 \pm 4.15$ &
$30.34 \pm 4.83$ &
$31.95 \pm 4.67$ &
$32.94 \pm 4.64$ &
$33.78 \pm 4.72$ &
$34.59 \pm 4.89$ &
$35.38 \pm 5.11$ \\
OPAL $B$ &
$25.98 \pm 2.05$ &
$31.43 \pm 2.30$ &
$32.34 \pm 2.07$ &
$32.72 \pm 1.87$ &
$33.04 \pm 1.74$ &
$33.39 \pm 1.69$ &
$33.79 \pm 1.72$ \\
DELPHI $B$ &
$23.76 \pm 2.09$ &
$29.42 \pm 2.68$ &
$31.02 \pm 2.71$ &
$32.10 \pm 2.73$ &
$33.05 \pm 2.77$ &
$33.96 \pm 2.82$ &
$34.84 \pm 2.88$ \\
\hline
\multicolumn{8}{|c|}{NNLO+NNLL logR CNO($f=2.0$)} \\
\hline
CLEO $D^{*+}$ &
$15.30 \pm 1.63$ &
$23.07 \pm 1.52$ &
$28.88 \pm 1.54$ &
$34.39 \pm 1.69$ &
$40.14 \pm 1.93$ &
$46.48 \pm 2.28$ &
$53.68 \pm 2.76$ \\
CLEO $D^{*0}$ &
$11.79 \pm 2.00$ &
$17.50 \pm 1.81$ &
$22.13 \pm 1.84$ &
$26.67 \pm 2.03$ &
$31.36 \pm 2.33$ &
$36.53 \pm 2.78$ &
$42.28 \pm 3.38$ \\
BELLE $D^{*+} \to D^0$ &
$14.50 \pm 0.89$ &
$22.30 \pm 0.96$ &
$28.67 \pm 1.05$ &
$35.00 \pm 1.20$ &
$41.86 \pm 1.42$ &
$49.66 \pm 1.74$ &
$58.82 \pm 2.19$ \\
BELLE $D^{*+} \to D^+$ &
$9.74 \pm 2.34$ &
$14.56 \pm 2.33$ &
$18.72 \pm 2.51$ &
$22.92 \pm 2.84$ &
$27.34 \pm 3.31$ &
$32.23 \pm 3.96$ &
$37.59 \pm 4.79$ \\
BELLE $D^{*0}$ &
$13.25 \pm 3.79$ &
$20.34 \pm 4.18$ &
$26.28 \pm 4.68$ &
$32.16 \pm 5.35$ &
$38.42 \pm 6.27$ &
$45.53 \pm 7.58$ &
$53.79 \pm 9.41$ \\
\hline
\end{tabular}
\caption{\label{tab:alpha}
Values of the parameter $\alpha$ of the non-perturbative fragmentation function $D^{np}_K(N)$, corresponding to the values of the first eight moments extracted from the experimental data and displayed in table~\ref{tab:DNP}.}
\end{minipage}
\end{sideways}
\end{table}

\subsection{Charm ratio}
\label{sec:charmratio}

We now consider an observable which is derived from experimental data,  but which is directly comparable to perturbative predictions, namely the ratio of Mellin moments of charmed hadrons data measured at LEP, at centre-of-mass energy $Q_f=91.2$~GeV, and at BELLE and CLEO, at $Q_i=10.6$~GeV,
\be
R_D(N,Q_i,Q_f) \equiv \frac{\sigma_D(N,Q_f=91.2~\text{GeV},m=1.5~\text{GeV},\mathrm{np~pars})}{\sigma_D(N,Q_i=10.6~\text{GeV},m=1.5~\text{GeV},\mathrm{np~pars})} \, .
\ee
As shown by the arguments in the equation above, the moments at the two energies $Q_i$ and $Q_f$ depend individually  on the mass of the heavy quark and on the parameters that describe the non-perturbative component that must be added to the perturbative part. However, these `low scale' dependencies drop out in the ratio\footnote{This is strictly true only in the non-singlet approximation. However, the inclusion of the singlet components has very small impact and does not change the picture. By the same token, the subtraction of the gluon-splitting contribution from the ALEPH data performed by the experimental collaboration is not expected to have a meaningful impact on this analysis. Furthermore, since the contribution of the initial conditions to this observable is negligible, the result is not affected by the Landau pole and its regularisation.}, leaving $R_D(N,Q_i,Q_f)$ to depend only on the two `high' scales $Q_i$ and $Q_f$. Since moments at these two scales are bridged by the DGLAP evolution, this ratio essentially measures the degree of correctness (or lack thereof) of the collinear factorisation picture that underlies the collinear resummation. Deviations from this picture, for instance under the form of power corrections larger than expected, can show up as a discrepancy between the experimental ratio and the perturbative prediction.

\begin{figure}[t]
\begin{center}
\includegraphics[width=12cm]{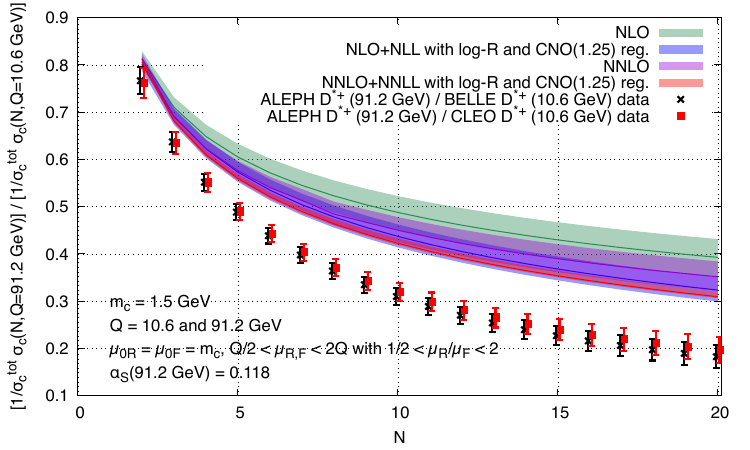}
\end{center}
\caption{\label{fig:charmratio} ALEPH/BELLE and ALEPH/CLEO ratios for $D^{*+}$ data, compared to theoretical predictions. The BELLE and CLEO data are corrected for initial state electromagnetic radiation effects~\cite{Cacciari:2005uk}.}
\end{figure}

This comparison had already been performed in~\cite{Cacciari:2005uk} using the NLO+NLL perturbative prediction for the fragmentation function and the $D^{*+}$ fragmentation data from ALEPH~\cite{ALEPH:1999syy} and from BELLE~\cite{Belle:2005mtx} (see footnote \ref{foot:belle}). A visible discrepancy with the data had been observed. We update that comparison to NNLO+NNLL in figure~\ref{fig:charmratio}. We see that the increased perturbative accuracy shrinks the uncertainty bands but, as expected and already mentioned in~\cite{Cacciari:2005uk}, it does not help in reconciling the prediction with the data, the discrepancy remaining largely the same. 

\section{Conclusions}

We have implemented in a C++ code, that will eventually be publicly released, the NNLO+NNLL calculation for heavy quark fragmentation in $e^+e^-$ collisions. The code can calculate charm and bottom production at $B$ factories and LEP energies, as well as top production at a hypothetical higher energy collider. A non-perturbative component, describing the fragmentation of the heavy quark into the heavy hadron, can be modeled with a simple function and convoluted with the perturbative result.

The first important observation that we make is that increasing the perturbative accuracy from NLO+NLL to NNLO+NNLL does not necessarily improve the quality or accuracy of the prediction. The need to regularise the soft-gluon resummation to avoid the non-perturbative Landau pole means that the result depends on this procedure, whose choice is not unique. We observe that, depending on the choice, the numerical value and the quality of convergence of the perturbative series can vary widely. 

Another observation is that, while the description of all heavy quarks (top, bottom and charm) is in principle perturbatively identical, important qualitative and quantitative differences can appear. 

Top production is usually very well behaved. Due to all scales involved being large, and the value of the strong coupling consequently small, one observes good convergence and limited dependence on the choice of Landau pole regularisation. Unfortunately, this is also the least phenomenologically relevant case.

Bottom production is quite dependent on the choice made for the Landau pole regularisation, but the fragmentation function calculated in perturbative QCD has perturbative uncertainties that are still under control, and it provides a sufficiently good baseline:  a simple model for the non-perturbative component is usually sufficient to provide a numerically small correction and achieve a reasonably good description of the data.

For charm production, due to the very low scales and large value of the coupling, all the sensitivities mentioned above are magnified, as one can expect. There are significant perturbative uncertainties, and significant quantitative differences between regularisation schemes. Moreover, in many cases the perturbative description provides a very poor description of the data, to the extent that a simple non-perturbative model cannot compensate sufficiently to describe them.

We conclude by observing that the time-honoured `perturbative QCD convoluted with a simple non-perturbative function', that has served us well up to NLO+NLL accuracy, appears to be challenged by the upgrade to NNLO+NNLL accuracy. The upgrade to a higher perturbative level leads to larger sensitivity to the details of the regularisation at the boundary with non-perturbative phenomena, largely washing out -- at the phenomenological level and especially for charm and bottom quarks -- any gain that one may have hoped to glean from the higher perturbative accuracy. We find it likely that the achievement of a really improved theoretical predictivity will necessitate a more thorough study and modeling of the interface between perturbative and non-perturbative regions.

\acknowledgments{We would like to thank Emanuele Bagnaschi, Gennaro Corcella, Thomas Gehrmann, Simone Marzani and Sven Moch for useful discussions during the course of this work. We are indebted to Sven Moch and Daniel De~Florian for providing us with the Form and Fortran implementations of the NNLO contribution to the $e^+e^-$ coefficient functions, and to Fabio Maltoni, Giovanni Ridolfi, Maria Ubiali and Marco Zaro for providing us with the Fortran implementation of the Mellin transforms of the initial conditions, and for discussions and  comparisons with our results. Discussions with Valerio Bertone on the usage of MELA are also gratefully acknowledged. We also thank the anonymous referee for many detailed and useful comments. LB's work has received funding from the Swiss National Science Foundation (SNF) under contract 200020-204200. The work of GS is partially supported by MIUR through the FARE grant R18ZRBEAFC.}

\appendix

\section{Perturbative Ingredients}

\subsection{Differential cross section}
\label{app:psdcs}

In the following we describe in detail the perturbative part of the differential cross section written in eq.~\eqref{eq:ff}, following the presentation of~\cite{Rijken:1996vr}. 
In Mellin space, the differential cross section for the production of a heavy quark $Q$ via $e^+e^-$ annihilation reads
\begin{align}\label{eq:sigQ}
    \sigma_Q(N,Q,m) = \sum_{i \in \{q,\bar q,g\}} \hat{\sigma}_{i}(N,Q,\muf) D_{i\to Q}(N,\muf,m) + {\cal O}\left(\left(\frac{m}{Q}\right)^p\right) \,\,  ,
\end{align}
where $p \geq 1$, and with the sum running over partonic channels.  Here and in the following we list in the arguments of the functions only the `real' scales on which they explicitly depend, and we  avoid writing also the `residual' scales, on which a quantity can still depend beyond a given perturbative order.

We can rewrite eq.~\eqref{eq:sigQ} as
\begin{align} \label{eq:sigSNS}
\sigma_{Q}(N,Q,m)&=\sum_{i\in\{q\}}\sigma^{(0)}_{i}\left(C^S_{q} D^S_{q\to Q}+C_{g} D_{g\to Q}+C^{NS}_{q}D^{NS}_{i\to Q}\right) \, ,
\end{align}
where the sum runs over $n_f$ quark flavours.
This form (obtained after summing over transverse and longitudinal components) makes it explicit the contribution of the leading order electroweak cross sections for production of a quark in $e^+e^-$ collisions, the $\sigma^{(0)}_{i}$, and that of the non-singlet, gluon and singlet QCD coefficient functions, $C^{NS}_{q}$, $C_{g}$ and $C^S_{q}$, described in Appendix \ref{app:pscf}.

The electroweak cross sections $\sigma^{(0)}_{i}$ are given by\footnote{As in~\cite{Cacciari:2005uk}, we complement the electroweak cross sections with a threshold factor for the heavy quarks and antiquarks,
\be
\sigma^{(0)}_{i}(Q) \to \sigma^{(0)}_{i}(Q)\left(1+\frac{2m_i^2}{Q^2}\right)\sqrt{1-\frac{4m_i^2}{Q^2}}\qquad i=c,b,t.
\ee
The numerical impact is however usually negligible at the combinations of heavy quark type and centre-of-mass energy $Q$ considered in this paper.
}
\begin{align} \label{eq:sigq}
\sigma^{(0)}_{i}(Q)=
\frac{4\pi\alpha^2}{3Q^2} N_C
\biggl[e^2_{\ell}e^2_{i}&+\frac{2Q^2\left(Q^2-M_Z^2\right)}{\left|Z(Q^2)\right|^2}e_{\ell}e_{i}C_{V,\ell}C_{V,i} \nonumber \\
&+\frac{(Q^2)^2}{\left|Z(Q^2)\right|^2}\left(C_{V,\ell}^2+C_{A,\ell}^2\right)\left(C_{V,i}^2+C_{A,i}^2\right)\biggr] \,  ,
\end{align}
In the above $Z(Q^2)=Q^2-M_Z^2+iM_Z\Gamma_Z$ with $M_Z$ and $\Gamma_Z$ mass and width of the $Z$-boson\footnote{We adopt the values of $M_Z=91.2 \,  \mathrm{GeV}$ and $\Gamma_Z=2.5  \, \mathrm{GeV}$ in our numerical implementation.} respectively and $e_{i}$ quark's $i$ electric charge.  The electroweak couplings of the charged lepton $\ell$ with charge $e_{\ell}=-1$ and quarks are 
\begin{align}
C_{A,\ell}&=\frac{1}{2\sin 2 \theta_W} \, , & C_{V,\ell}=&-C_{A,\ell}(1-4\sin ^2 \theta_W) \,  , \nonumber \\
C_{A,u}&=-C_{A,d}=-C_{A,\ell} \,  , \nonumber \\
C_{V,u}&=C_{A,\ell}\left(1-\frac{8}{3}\sin^2\theta_W \right) \, , &  C_{V,d}=&-C_{A,\ell}\left(1-\frac{4}{3}\sin ^2 \theta_W\right) \,,
\end{align}
with $\theta_W$ Weinberg angle and $\sin ^2 \theta_W = 0.23 $.

We also introduce the total cross section for production of a heavy quark $Q=c,b,t$,
\begin{align}\label{eq:sigtot}
	&\sigma^{tot}_{Q}(Q)=\sigma^{(0)}_{Q}(Q)\left\{1+\frac{\aS(\mur)}{2\pi}\frac{3}{2}C_F +\left(\frac{\aS(\mur)}{2\pi}\right)^2\left[-\frac{3}{8}C_F^2 \right. \right.\nonumber \\
	&\left.\left.+C_A C_F\left(-\frac{11}{4}\ln\frac{Q^2}{\mur^2}-11\zeta(3)+\frac{123}{8}\right)+n_f C_F T_R\left(\ln\frac{Q^2}{\mur^2}+4\zeta(3)-\frac{11}{2}\right)\right]+\mcO\left(\as^3\right)\right\} \, ,
\end{align}
with $n_f$ number of active flavours (including the heavy quark) and QCD colour factors $C_F=4/3$, $C_A=3$, $T_R=1/2$.

The $D_{i\to Q}$ functions in eq.~\eqref{eq:sigQ} are the \textit{perturbative} fragmentation functions at the hard scale of the process, namely initial conditions $D_{j\to Q}(\muzf,m)$ evolved through DGLAP equations from the initial factorisation scale $\muzf\sim m$ to the final one  $\muf\sim Q$,
\begin{equation}
    D_{i\to Q}(N,\muf,m)=\sum_j E_{ij}(N,\muf,\muzf )D_{j\to Q}(N,\muzf,m) \,  .
\end{equation}
The initial conditions and their soft-gluon resummation are described in detail in Appendix \ref{app:psic}. Their DGLAP evolution is performed using the MELA package as described in Appendix~\ref{app:melasetup}.

The relations between the partonic initial conditions $D_{i\to Q}$ and the singlet and non-singlet combinations that enter eq.~(\ref{eq:sigSNS}) are\footnote{Since we are inclusive over the angle of emission of the identified hadron with respect to the electron beam direction in the center-of-mass frame, we have no contribution from the asymmetric component defined in~\cite{Rijken:1996vr}.}
\begin{align} \label{eq:SNSFF}
    D_{q\to Q}^{S}&=\frac{1}{n_f}\sum_{i\in\{q\}}\left(D_{i\to Q}+D_{\bar{i}\to Q}\right) \, ,  \\
    D^{NS}_{i\to Q}&=D_{i\to Q}+D_{\bar{i}\to Q}-D^{S}_{q\to Q} \, .
\end{align}

Finally, we can rewrite eq.~\eqref{eq:sigSNS} in a partonic basis (i.e.\ recover the form of eq.~(\ref{eq:sigQ})) as
\begin{align}
\sigma_{Q}(N,Q,m)= \left(\sum_{j\in\{q\}}\sigma^{(0)}_{j}\right) \left[C_{g} D_{g\to Q}+\sum_{i\in\{q\}}(C_{i}  D_{i\to Q}+C_{\bar{i}} D_{\bar{i}\to Q})\right] \, ,
\end{align}
with
\begin{align}\label{eq:cqi}
C_{i}=\frac{1}{n_f} \left(C_{q}^{S}-C_{q}^{NS}\right)+ \frac{\sigma^{(0)}_{i}}{\sum_{j\in\{q\}}\sigma^{(0)}_{j}} C_{q}^{NS} \, .
\end{align}
At NLO the singlet and the non-singlet coefficient function are equal and the first term cancels. At NNLO the difference $C_{q}^{S}-C_{q}^{NS}$ is the pure-singlet (PS) contribution, which vanishes in the large-$N$ limit. In the partonic-channel basis, the quarks coefficient functions contain the quark's charge, leading to a distinction between (anti)up-type and (anti)down-type.

The coefficient functions and the soft-gluon resummation of the $C_i$ in eq.~(\ref{eq:cqi}) are described in detail in Appendix \ref{app:pscf}.

\subsection{Perturbative parameters}
\label{app:softingr}

We list here the parameters that enter the perturbative results and the soft-gluon resummation up to NNLL accuracy.

The strong coupling $\aS(\mu)$ evolves as
\be
\frac{d\aS}{d\ln\mu^2} = \beta(\aS) \, ,
\ee
and the coefficients of the QCD $\beta$-function 
\begin{equation}
    \beta(\aS)=-b_0 \aS^2-b_1\aS^3-b_2\aS^4+\mcO\left(\aS^5\right) \, , 
\end{equation}
in a theory with $n$ massless flavours read
\begingroup
\allowdisplaybreaks
\begin{align}
    b_0^{(n)}&=\frac{11C_A-4T_Rn}{12\pi} \, , \nonumber\\
    b_1^{(n)}&=\frac{17C_A^2-(10C_AT_R-6C_FT_R)n}{24\pi^2} \, , \nonumber\\
    b_2^{(n)}&=\frac{1}{(4\pi)^3}\left[\frac{2857}{54}C_A^3-\left(\frac{1415}{54}C_A^2+\frac{205}{18}C_AC_F-C_F^2\right)n+\left(\frac{79}{54}C_A+\frac{11}{9}C_F\right)n^2\right]\, ,
\end{align}
\endgroup
with $C_F = 4/3$, $C_A = 3$, $T_R = 1/2$. In practice, we use the evolution of the strong coupling $\as(\mu$) as implemented in MELA, see Appendix~\ref{app:melasetup}.

We define
\be
A[\as] = \sum_{k=0}^\infty \left(\frac{\as}{\pi}\right)^k A_k, \qquad
B[\as] = \sum_{k=0}^\infty \left(\frac{\as}{\pi}\right)^k B_k, \qquad
H[\as] = \sum_{k=0}^\infty \left(\frac{\as}{\pi}\right)^k H_k\, .
\ee
The coefficients of the cusp anomalous dimensions $A[\as]$ up to $\mathcal{O}(\as^3)$ and of the functions $B[\as]$ and $H[\as]$ up to $\mathcal{O}(\as^2)$~\cite{Moch:2009my,Cacciari:2001cw,Aglietti:2006yf,Maltoni:2022bpy} are
\begingroup
\allowdisplaybreaks
\begin{align}
  	A_1&=C_F \, , \nonumber \\
    A_2^{(n)}&=\frac{1}{2}C_F\left(C_A\left(\frac{67}{18}-\zeta(2)\right)-\frac{5}{9}n\right)\, , \nonumber \\
    A_3^{(n)}&=C_F\left[C_A^2\left(\frac{245}{96}+\frac{11}{24}\zeta(3)-\frac{67}{36}\zeta(2)+\frac{11}{8}\zeta(4)\right)\right. \nonumber \\
    &\quad \quad \quad -C_An\left(\frac{209}{432}+\frac{7}{12}\zeta(3)-\frac{5}{18}\zeta(2)\right)-\left.C_F n\left(\frac{55}{96}-\frac{\zeta(3)}{2}\right)-\frac{n^2}{108}\right] \, ,  \nonumber \\
    B_1&= -\frac{3}{2}C_F \, , \nonumber \\
    B_2^{(n)}&= 2 C_F\left[C_A\left(-\frac{3155}{864}+\frac{11}{12}\zeta(2)+\frac{5}{2}\zeta(3)\right) \right.  \nonumber \\
    &\left. \quad \quad \quad -C_F\left(\frac{3}{32}+\frac{3}{2}\zeta(3)-\frac{3}{4}\zeta(2)\right) + n\left(\frac{247}{432}-\frac{\zeta(2)}{6}\right) \right]\, , \nonumber \\
    H_1&=-C_F \, , \nonumber \\
    H_2^{(n)}&=-C_F\left[\frac{\pi b_0^{(n)}}{9}+C_A\left(\frac{9}{4}\zeta(3)-\frac{\pi^2}{12}-\frac{11}{18}\right)\right]  \nonumber \\
     &\quad \quad \quad
    +\frac{C_FT_R}{54}\left(9\ln^2\frac{m^2}{\muzf^2}+30\ln\frac{m^2}{\muzf^2}+28\right) \,  .
\end{align}
\endgroup
For the resummation of the coefficient functions up to NNLL we use
\begin{align}
\{b_0^{(n_f)},b_1^{(n_f)},b_2^{(n_f)},A_1,A_2^{(n_f)},A_3^{(n_f)},B_1,B_2^{(n_f)}\} \, ,
\end{align}
and for the resummation of the initial conditions up to NNLL
\begin{align}
\{b_0^{(n_l)},b_1^{(n_l)},b_2^{(n_l)},A_1,A_2^{(n_l)},A_3^{(n_l)},H_1,H_2^{(n_l)}\} \, ,
\end{align}
with $n_l = n_f-1$.

\subsection{QCD corrections to the coefficient functions}
\label{app:pscf}

The coefficient functions are process-dependent objects,  computed in perturbative QCD with $n_f$ massless partons. In this section, it is understood that $\aS = \aS^{(n_f)}$. The (sum of transverse and longitudinal) singlet and non-singlet combinations up to $\mcO(\as^2)$ read
\begin{align} \label{CqCg}
C_{q}^{NS}(N,Q,\muf)&=1+\frac{\aS(\mur)}{2\pi}\hat{a}_{q}^{(1)}(N,Q,\muf)\nonumber\\
+&\left(\frac{\aS(\mur)}{2\pi}\right)^2\frac{1}{4}\left(\hat{a}_{q}^{(2),NS}(N,Q,\muf)-8\pi b_0 \ln\frac{\muf^2}{\mur^2}\hat{a}_{q}^{(1)}(N,Q,\muf)\right)+\mcO \left(\aS^3\right) \, , \nonumber \\
C_{g}(N,Q,\muf)&=\frac{\aS(\mur)}{2\pi}\hat{a}_{g}^{(1)}(N,Q,\muf)\nonumber \\
+&\left(\frac{\aS(\mur)}{2\pi}\right)^2\frac{1}{4}\left(\hat{a}_{g}^{(2),S}(N,Q,\muf)-8\pi  b_0 \ln\frac{\muf^2}{\mur^2}\hat{a}_{g}^{(1)}(N,Q,\muf)\right) +\mcO \left(\aS^3\right) \, ,  \nonumber \\
C_{q}^{PS}(N,Q,\muf)&=\left(\frac{\aS(\mur)}{2\pi}\right)^2\frac{1}{4}\hat{a}_{q}^{(2),PS}(N,Q,\muf)+\mcO \left(\aS^3\right) \, , \nonumber \\
C_{q}^{S}(N,Q,\muf)&=C_{q}^{NS}(N,Q,\muf)+C_{q}^{PS}(N,Q,\muf) \,  .
\end{align}
Since the two basis described in Appendix \ref{app:psdcs} are degenerate at NLO, the $\hat{a}^{(1)}_i$ coefficients are insensitive to the basis chosen.
The coefficients $\hat{a}$ are given e.g.\ in~\cite{Mele:1990cw,Nason:1997nu} at NLO and in~\cite{Rijken:1996vr} ($x$-space),~\cite{Blumlein:2006rr,Mitov:2006wy} ($N$-space) at NNLO.  The factors $1/4$ are meant to make contact with the notation of~\cite{Blumlein:2006rr}, where the expansion is given in terms of $a_s=\aS/(4\pi)$, and whose $\hat{a}^{(2)}$ expressions are used in the Fortran implementation that we integrate into our code.
The terms proportional to $\aS^2\ln (\muf^2/\mur^2)$ in $C_q^{NS}$ and $C_g^{S}$ account for the $\mur \neq \muf$ scale-dependence from the NLO running of the strong coupling, 
$\as(\muf) =  \as(\mur)\left(1-b_0 \as \ln (\muf^2/\mur^2)\right) + {\cal O}(\as^3)$, with $b_0=(11C_A-4T_R n_f)/(12\pi)$. 

The NLO coefficients in $e^+e^-$ collisions read
\begingroup
\allowdisplaybreaks
\begin{align}
    \hat{a}_{q}^{(1)}(N,Q,\muf)=C_F&\left[ \ln\frac{Q^2}{\muf^2}\left(\frac{3}{2}+\frac{1}{N(N+1)}-2S_1(N)\right)\right. +\frac{1}{N}-2\frac{2N+1}{N^2(N+1)^2}  \nonumber\\
   & -4\psi_1(N+1)+S_1^2(N)-\frac{1}{N(N+1)}S_1(N)+\frac{1}{(N+1)^2}+S_2(N) 
   \nonumber\\
    &\left. +\frac{3}{2}S_1(N)-\frac{3}{2}\frac{1}{N+1}+\frac{2}{3}\pi^2-\frac{9}{2}\right] \, , \\
    \hat{a}_{g}^{(1)}(N,Q,\muf)=C_F&\left[ \frac{2(2+N+N^2)}{N(N^2-1)}\ln\frac{Q^2}{\muf^2}+4\left(-\frac{2}{(N-1)^2}+\frac{2}{N^2}-\frac{1}{(N+1)^2}\right)\right. \nonumber \\
    & \left.-2 \left(\frac{2}{N-1}S_1(N-1)-\frac{2}{N}S_1(N)+\frac{1}{N+1}S_1(N+1)\right) \right] \,  ,
\end{align}
\endgroup
with the special functions $S_i(N)$, 
\begin{align}
    S_1(N)&=\psi_0(N+1)-\psi_0(1) \, , \nonumber \\
    S_2(N)&=-\psi_1(N+1)+\psi_1(1) \, , \nonumber \\
    S_3(N)&=\frac{1}{2}\left(\psi_2(N+1)-\psi_2(1)\right) \, ,
\end{align}
written in terms of the polygamma functions $\psi_m(N)$.

To account for soft-gluon effects, the quark coefficient function $C_i$ of eq.~\eqref{eq:cqi}\footnote{In practice all the large-$N$ logarithmic behaviour is found in the NS quark coefficient function, since the difference between singlet and non singlet contribution in eq.~\eqref{eq:cqi} vanishes in the large-$N$ limit.} must be resummed to all orders in $\as$.  
The Sudakov-resummed coefficient function, up to NNLL accuracy is
\begin{align}
\label{eq:cfres}
C_{q}^{res}(N,Q,\muf) =\Biggl(1&+\frac{\as(\mur)}{2\pi}\bar{a}^{(1)}_{q}(Q,\muf)
\nonumber\\
&+\left(\frac{\as(\mur)}{2\pi}\right)^2\bar{a}^{(2)}_{q}(Q,\muf,\mur) + \mcO\left(\as^3\right) \Biggr) C^{sud}_q(N,Q,\muf,\mur) \, ,
\end{align}
where $\bar{a}^{(1)}_{q}(Q,\muf)$ and $\bar{a}^{(2)}_{q}(Q,\muf)$ are the constant (i.e.\ $N$-independent) terms of the large-$N$ limit of $\hat{a}^{(1)}_{q}(N,Q,\muf)$ and $\hat{a}^{(2),NS}_{q}(N,Q,\muf,\mur)$ respectively. They read
\begingroup
\allowdisplaybreaks
  \begin{align}
    \label{eq:cfconst1}
\bar{a}^{(1)}_{q}&(Q,\muf)=2C_F\left[\frac{5}{12}\pi^2-\frac{9}{4}+\frac{1}{2}\gamma_E^2+\frac{3}{4}\gamma_E+\left(\frac{3}{4}-\gamma_E\right)\ln\frac{Q^2}{\muf^2}\right] \, , \\
\label{eq:cfconst2}
    \bar{a}^{(2)}_{q}&(Q,\muf,\mur)\nonumber\\
      &=\frac{C_F C_A}{4} \biggl[\frac{22 \gamma_E^3}{9}-\frac{11}{3}
       \log \biggl(\frac{\muf^2}{\mur^2}\biggr) \biggl(2
       \gamma_E^2+(3-4 \gamma_E) \log
       \biggl(\frac{Q^2}{\muf^2}\biggr)+3 \gamma_E+10
       \zeta(2)-9\biggr)\nonumber\\
       &\quad +\frac{1}{18} \biggl(-132
       \gamma_E^2-\gamma_E (734-144 \zeta(2))-396 \zeta(2)-216
       \zeta(3)+645\biggr) \log \biggl(\frac{Q^2}{\muf^2}\biggr) \nonumber \\
       &\quad-\frac{2 \pi ^2
       \gamma_E^2}{3}+\frac{367 \gamma_E^2}{18}+\frac{11}{6} (4
       \gamma_E-3) \log ^2\biggl(\frac{Q^2}{\muf^2}\biggr)-40
       \gamma_E \zeta(3)-\frac{11 \pi ^2 \gamma_E}{9}+\frac{3155
       \gamma_E}{54}\nonumber\\
       &\quad  +\frac{464 \zeta(3)}{9}
       -\frac{23 \pi ^4}{60}+\frac{1657
       \pi ^2}{108}-\frac{5465}{72}\biggr]\nonumber\\
     &+\frac{C_F^2}{4} \biggl[2 \gamma_E^4+6
       \gamma_E^3+\frac{10 \pi ^2 \gamma_E^2}{3}-\frac{27
       \gamma_E^2}{2}\nonumber\\
       &\quad +\frac{1}{2} \biggl(-16 \gamma_E^3-12
       \gamma_E^2-80 \gamma_E \zeta(2)+90 \gamma_E+36
       \zeta(2)+48 \zeta(3)-51\biggr) \log
       \biggl(\frac{Q^2}{\muf^2}\biggr)\nonumber\\
       &\quad +\frac{1}{2} (4 \gamma_E-3)^2
       \log ^2\biggl(\frac{Q^2}{\muf^2}\biggr)+24 \gamma_E \zeta(3)+3
       \pi ^2 \gamma_E \nonumber\\
       &\quad -\frac{51 \gamma_E}{2}-66 \zeta(3)+\frac{61
       \pi ^4}{45}-\frac{35 \pi ^2}{4}+\frac{331}{8}\biggr]\nonumber\\
      &+ \frac{C_F n_f}{4} \biggl[-\frac{4
       \gamma_E^3}{9}+\frac{2}{3} \log
       \biggl(\frac{\muf^2}{\mur^2}\biggr) \biggl(2 \gamma_E^2+(3-4
       \gamma_E) \log \biggl(\frac{Q^2}{\muf^2}\biggr)+3
       \gamma_E+10 \zeta(2)-9\biggr)\nonumber\\
       &\quad +\frac{1}{9} \biggl(12
       \gamma_E^2+58 \gamma_E+36 \zeta(2)-57\biggr) \log
       \biggl(\frac{Q^2}{\muf^2}\biggr)-\frac{29
       \gamma_E^2}{9}+\frac{1}{3} (3-4 \gamma_E) \log
       ^2\biggl(\frac{Q^2}{\muf^2}\biggr)\nonumber\\
       &\quad +\frac{2 \pi ^2
       \gamma_E}{9}-\frac{247 \gamma_E}{27}+\frac{4
       \zeta(3)}{9}-\frac{143 \pi ^2}{54}+\frac{457}{36}\biggr]
    \end{align}
\endgroup
We had extracted the constant $\bar a^{(2)}_{q}$ from the non-singlet coefficient function presented in~\cite{Mitov:2006wy}, and we have found it to coincide with the one successively published in~\cite{Czakon:2022pyz}.

The Sudakov factor $C^{sud}_q$ takes the exponential form~\cite{Cacciari:2001cw,Moch:2009my,Czakon:2022pyz}
\begin{align}\label{Csud}
&\ln C^{sud}_q(N,Q,\muf) \nonumber \\
&= \int_0^1 dz \frac{z^{N-1}-1}{1-z}\left[\int_{\muf^2}^{Q^2(1-z)} \frac{d\mu^2}{\mu^2}\, A[\as(\mu)] + \frac{1}{2} B[\as(Q\sqrt{1-z})] + D[\as(Q(1-z))]\right]\nonumber \\
&=\ln Ng_1(\lambda)+g_2(\lambda;Q,\muf,\mur)+\as(\mur) g_3(\lambda;Q,\muf,\mur) + {\cal O}(\as^k \ln^{k-2}N)\, ,
\end{align}
where we have defined $\lambda=\as(\mur) b_0 \ln N$ and the functions $A[\as]$ and $B[\as]$ are given in Appendix~\ref{app:softingr}. The function $D[\as]$ can be neglected at NNLL accuracy~\cite{Cacciari:2001cw,Moch:2009my}.

The functions $g_k$ up to at least $k=3$ are available in the literature~\cite{Cacciari:2001cw,Moch:2009my,Czakon:2022pyz}. We have recalculated them and found perfect agreement.  They read
\begingroup
\allowdisplaybreaks
\begin{align}
    g_1(\lambda)&= \frac{A_1}{\pi b_0 \lambda}[\lambda+(1-\lambda)\ln(1-\lambda)] \,  ,  \\
    g_2(\lambda;Q,\muf,\mur)&= \frac{A_1 b_1}{\pi b_0^3}\left(\lambda+\ln(1-\lambda)+\frac{1}{2}\ln^2(1-\lambda)\right)+\frac{B_1 -2A_1\gamma_E}{2\pi b_0}\ln(1-\lambda) \nonumber \\
    &-\frac{1}{\pi b_0}(\lambda+\ln (1-\lambda)) \left(\frac{A_2 }{\pi b_0}-A_1\ln \frac{Q^2}{\mur^2}\right)-\frac{A_1}{\pi b_0}\lambda\ln\frac{Q^2}{\muf^2} \, ,  \\
    g_3(\lambda;Q,\muf,\mur)&=\frac{A_1}{2\pi}(\zeta(2)+\gamma_E^2)\frac{\lambda}{1-\lambda}-\frac{A_1\gamma_E b_1}{\pi b_0^2 (1-\lambda)}(\lambda+\ln(1-\lambda))\nonumber \\
  &+\frac{A_1 b_1^2}{\pi b_0^4(1-\lambda)}\left(\frac{\lambda^2}{2}+\lambda\ln(1-\lambda)+\frac{1}{2}\ln^2(1-\lambda)\right)\nonumber \\
  &+\frac{A_1 b_2}{\pi b_0^3}\left(\frac{\lambda}{1-\lambda}+\ln(1-\lambda)-\frac{\lambda^2}{2(1-\lambda)}\right)+\frac{A_3 }{\pi^3b_0^2}\frac{\lambda^2}{2(1-\lambda)}\nonumber \\
  &+\frac{A_2 \gamma_E}{\pi^2 b_0}\frac{\lambda}{1-\lambda}-\frac{A_2 b_1}{\pi^2 b_0^3}\frac{1}{1-\lambda}\left(\lambda+\ln(1-\lambda)+\frac{\lambda^2}{2}\right) \nonumber \\ 
  &-\frac{B_2 }{2\pi^2 b_0}\frac{\lambda}{1-\lambda}+\frac{B_1 b_1}{2\pi b_0^2(1-\lambda)}\left(\lambda+\ln (1-\lambda)\right)-\frac{B_1 \gamma_E}{2\pi}\frac{\lambda}{1-\lambda} \nonumber \\
  &-\frac{A_2 }{\pi^2b_0}\lambda \ln \frac{Q^2}{\muf^2}-\frac{A_1}{2\pi}\lambda\ln^2\frac{Q^2}{\muf^2}+\frac{A_1}{\pi}\lambda\ln\frac{Q^2}{\mur^2}\ln\frac{Q^2}{\muf^2} \nonumber \\
	&+\frac{1}{1-\lambda}\left[\frac{A_1 b_1}{\pi b_0^2}\left(\lambda+\ln(1-\lambda)\right)-\frac{A_1\gamma_E}{\pi}\lambda-\frac{A_2 }{\pi^2 b_0}\lambda^2\right]\ln\frac{Q^2}{\mur^2} \nonumber \\
	&+\frac{A_1}{2\pi}\frac{\lambda^2}{1-\lambda}\ln^2\frac{Q^2}{\mur^2} +\frac{B_1 }{2\pi}\frac{\lambda}{1-\lambda}\ln\frac{Q^2}{\mur^2} \, .
\end{align}
\endgroup

For completeness, we give here the expansion in $\as$ of the Sudakov factor up to $\mcO\left(\as^2\right)$, to be used in the matching of the resummed result to the fixed order one:
\begin{align}
[C^{sud}_q(N,Q,\muf)]_{\as^2}=1+\frac{\as(\mur)}{2\pi} \left.C^{sud}_q(N,Q,\muf)\right|_{\as} +\left(\frac{\as(\mur)}{2\pi}\right)^2 \left.C^{sud}_q(N,Q,\muf,\mur)\right|_{\as^2} \, ,
\end{align}
with
\begin{align}
&\left.C^{sud}_q(N,Q,\muf)\right|_{\as}=\left[-B_1 +2A_1\left(\gamma_E-\ln\frac{Q^2}{\muf^2}\right)\right]\ln N+A_1\ln^2 N  \, , \\
&\left.C^{sud}_q(N,Q,\muf,\mur)\right|_{\as^2} = 2\biggl[-B_2 +2A_2 \gamma_E - B_1 b_0\gamma_E\pi+A_1 b_0\pi(\gamma_E^2+\zeta(2))\nonumber \\
&\quad \quad -2A_2 \ln\frac{Q^2}{\muf^2} +\left(B_1 b_0\pi-2A_1 b_0\gamma_E\pi\right)\ln\frac{Q^2}{\mur^2}+2A_1 b_0 \pi \ln\frac{Q^2}{\mur^2}\ln\frac{Q^2}{\muf^2} \nonumber \\
&\quad \quad -A_1 b_0\pi\ln^2\frac{Q^2}{\muf^2} \biggr]\ln N -2\left[-A_2 +A_1 b_0 \pi \ln \frac{Q^2}{\mur^2}+\frac{\pi}{2}\left(B_1 b_0-2A_1 b_0\gamma_E\right)\right]\ln^2 N \nonumber \\
&\quad \quad +\frac{2}{3}\pi A_1 b_0\ln^3N +\frac{1}{2}\left(-B_1 \ln N+2A_1 \gamma_E \ln N +A_1 \ln^2 N-2A_1 \ln\frac{Q^2}{\muf^2}\ln N\right)^2 . 
\end{align}

\subsection{Initial conditions}
\label{app:psic}

As first introduced in~\cite{Mele:1990cw}, the initial conditions $D_{i\to Q}$ for heavy hadron production are perturbative objects, and therefore admit a perturbative expansion in $\aS$. The $\mcO(\as^2)$ corrections were computed in~\cite{Melnikov:2004bm,Mitov:2004du} in $x$-space and more recently transformed to $N$-space~\cite{Maltoni:2022bpy}. 
The calculations of~\cite{Melnikov:2004bm,Mitov:2004du} were performed in an \MSbar renormalisation scheme for ultraviolet divergences where both massless and massive flavours contribute to the evolution of the running coupling (i.e.\ an $n_f$-scheme).  To match the fixed order results with the resummed one, where only massless flavours contribute  the evolution of the running coupling (an $n_l$ scheme), the initial conditions must be written in the $n_l$-scheme. 
In $N$-space they read, expanded in powers of $\as^{(n_l)}$, i.e.\ renormalised with $n_l = n_f-1$ massless flavours,
\begingroup
\allowdisplaybreaks
\begin{align}
    D_{Q\to Q}(N,\muzf,m)&=1+\frac{\aS^{(n_l)}(\muzr)}{2\pi}d^{(1)}_{Q\to Q}(N,\muzf,m)+ \left(\frac{\aS^{(n_l)}(\muzr)}{2\pi}\right)^2\Bigg[d^{(2)}_{Q\to Q}(N,\muzf,m) \nonumber \\
&-\left(2\pi b_0^{(n_l+1)}\ln\frac{\muzf^2}{\muzr^2}+\frac{2}{3}T_R\ln\frac{m^2}{\muzr^2}\right)d^{(1)}_{Q\to Q}(N,\muzf,m) \Bigg]+ \mcO\left(\aS^3\right) \, ,\nonumber \\
    D_{\bar{Q}\to Q}(N,\muzf,m) &= \left(\frac{\aS^{(n_l)}(\muzr)}{2\pi}\right)^2d^{(2)}_{\bar{Q}\to Q}(N,\muzf,m)+ \mcO\left(\aS^3\right) \, ,\nonumber \\
    D_{g\to Q}(N,\muzf,m)&=\frac{\aS^{(n_l)}(\muzr)}{2\pi}d^{(1)}_{g\to Q}(\muzf,m)+ \left(\frac{\aS^{(n_l)}(\muzr)}{2\pi}\right)^2\Bigg[d^{(2)}_{g\to Q}(N,\muzf,m) \nonumber \\
&-\left(2\pi b_0^{(n_l+1)}\ln\frac{\muzf^2}{\muzr^2}+\frac{2}{3}T_R\ln\frac{m^2}{\muzr^2}\right)d^{(1)}_{g\to Q}(N,\muzf,m) \Bigg]+\mcO\left(\aS^3\right) \, , \nonumber \\
    D_{q\to Q}(N,\muzf,m)&= D_{\bar{q}\to Q}(N,\muzf,m) = \left(\frac{\aS^{(n_l)}(\muzr)}{2\pi}\right)^2d^{(2)}_{q\to Q}(N,\muzf,m) +\mcO\left(\aS^3\right) \, , \label{eq:dini}
\end{align}
\endgroup
where $Q = c,b$ or $t$, and $q$ are the flavours lighter than the quark identified as heavy, and therefore considered massless.
The terms proportional to $\as^2\ln (\muzf^2/\muzr^2)$ account for the renormalisation-scale dependence at NNLO from the running of the coupling at NLO.  
The terms proportional to $\as^2\ln (m^2/\muzr^2)$  in eq.~\eqref{eq:dini} account for the change of scheme, as described in~\cite{Maltoni:2022bpy}.  For sake of clarity, in eq.~\eqref{eq:dini} $b_0^{(n_l+1)}$ shows explicitly the number of flavours used in its definition.\footnote{This was not an issue for the coefficient functions, as the fixed order part is computed in the same \textit{massless} $n_f$-scheme as the resummation.}

The fixed order NLO coefficients, see e.g.\ \cite{Mele:1990cw,Nason:1997nu}, are given by
\begin{align}
    d^{(1)}_{Q\to Q}(N,\muzf,m) &= C_F\left[\ln\frac{\muzf^2}{m^2}\left(\frac{3}{2}+\frac{1}{N(N+1)}-2S_1(N)\right)  -2S_1^2(N)+\frac{2}{N(N+1)}S_1(N)\right. \nonumber\\
    & \left. \quad \quad \quad -\frac{2}{(N+1)^2}-2S_2(N)+2-\frac{1}{N(N+1)}+2S_1(N)\right] \,  ,\\
    d^{(1)}_{g\to Q}(N,\muzf,m) &= T_R\left(\frac{N^2+N+2}{N(N+1)(N+2)}\right)\ln\frac{\muzf^2}{m^2} \, ,
\end{align}
while the NNLO coefficients $d^{(2)}_{i\to Q}$ can be found in refs.~\cite{Melnikov:2004bm,Mitov:2004du}.

To account for soft-gluons effects, the heavy quark initial condition $D_{Q\to Q}$ must be resummed to all orders in $\as$.
The Sudakov-resummed heavy quark initial condition reads
\begin{align}
\label{eq:inires}
    &D^{res}_{Q\to Q}(N,\muzf,m)=\Biggl(1+\frac{\as^{(n_l)}(\muzr)}{2\pi} \bar{d}^{(1)}_{Q\to Q}(\muzf,m)+\left(\frac{\as^{(n_l)}(\muzr)}{2\pi}\right)^2 \biggl(\bar{d}^{(2)}_{Q\to Q}(\muzf,m) \nonumber \\
  & \quad - \left(2\pi b_0^{(n_l+1)}\ln\frac{\muzf^2}{\muzr^2}+\frac{2}{3}T_R\ln\frac{m^2}{\muzr^2}\biggr)\bar{d}^{(1)}_{Q\to Q}(\muzf,m)  \right)+\mcO(\as^3)\Biggr) D^{sud}_{Q\to Q}(N,\muzf,m) \, .
\end{align}
The constants $\bar{d}^{(1)}_{Q\to Q}(\muzf,m) $ and $\bar{d}^{(2)}_{Q\to Q}(\muzf,m) $ from the large-$N$ limits of $d^{(1)}_{Q\to Q}(N,\muzf,m)$ and $d^{(2)}_{Q\to Q}(N,\muzf,m)$ are given in~\cite{Cacciari:2001cw,Melnikov:2004bm}. We copy them here as written in~\cite{Maltoni:2022bpy} for completeness:
\be
\label{eq:iniconst1}
\bar{d}^{(1)}_{Q\to Q}(\muzf,m)  = -C_F\left[2\gamma_E^2+\frac{3}{2}\ln\frac{m^2}{\muzf^2} -2\gamma_E\left(1+\ln\frac{m^2}{\muzf^2}\right)-2\left(1-\frac{\pi^2}{6}\right)\right]
\ee
and
\begingroup
\allowdisplaybreaks
\begin{align}
  \label{eq:iniconst2}
&\bar{d}^{(2)}_{Q\to Q}(\muzf,m) \nonumber \\
&=C_A C_F \Bigg[\left(\frac{11}{8}-\frac{11 \gamma_E }{6}\right) \log^2\frac{m^2}{ \muzf^2}
  -\left(-3 \zeta (3)+\frac{35}{8}-\frac{34 \gamma_E}{9}-\frac{11 \gamma_E ^2}{3}+\frac{\gamma_E  \pi ^2}{3}\right)\log\frac{m^2}{ \muzf^2} 
  \nonumber\\
  &\quad +9 \gamma_E  \zeta (3)-\frac{97 \zeta (3)}{18}+\frac{\pi ^4}{12}+\frac{\gamma_E ^2 \pi
     ^2}{3}-\frac{14 \gamma_E  \pi ^2}{9}+\frac{7 \pi ^2}{54}-\frac{22 \gamma_E ^3}{9}-\frac{34 \gamma_E ^2}{9}-\frac{55 \gamma_E }{27}\nonumber\\
  &\quad +\frac{1141}{288}+\pi^2 \log 2\Bigg]
     \nonumber\\
     &+C_F^2 \Bigg[\frac{1}{8} (4 \gamma_E -3)^2 \log^2\frac{m^2}{ \muzf^2}
     -\left(6 \zeta (3)+\frac{27}{8}-\gamma_E -7 \gamma_E ^2+4\gamma_E ^3-\pi ^2+\frac{2 \gamma_E  \pi ^2}{3}\right)\log\frac{m^2}{ \muzf^2} 
     \nonumber\\
     &\quad-\frac{3 \zeta (3)}{2}-\frac{11 \pi ^4}{180}+\frac{2 \gamma_E ^2 \pi ^2}{3}-\frac{2 \gamma_E  \pi^2}{3}+\frac{\pi ^2}{4}+2 \gamma_E ^4-4 \gamma_E ^3-2 \gamma_E ^2+4 \gamma_E +\frac{241}{32}-2 \pi ^2 \log 2\Bigg]
     \nonumber\\
     &+C_F T_R
     \Bigg[\left(\frac{2 \gamma_E }{3}-\frac{1}{2}\right) \log^2\frac{m^2}{ \muzf^2}-\left(-\frac{3}{2}+\frac{8 \gamma_E }{9}+\frac{4 \gamma_E ^2}{3}\right)
     \log\frac{m^2}{ \muzf^2}+\frac{2 \zeta (3)}{3}-\frac{\pi ^2}{3}\nonumber\\
     &\quad -\frac{56 \gamma_E }{27}+\frac{3139}{648}\Bigg]
     \nonumber\\
     &+C_F T_R n_l
     \Bigg[\left(\frac{2 \gamma_E }{3}-\frac{1}{2}\right) \log^2\frac{m^2}{ \muzf^2}-\left(-\frac{3}{2}+\frac{8 \gamma_E }{9}+\frac{4 \gamma_E ^2}{3}\right)
     \log\frac{m^2}{ \muzf^2}
     \nonumber\\
     &\quad
     -\frac{2 \zeta (3)}{9}+\gamma  \left(\frac{4 \pi ^2}{9}-\frac{4}{27}\right)-\frac{4 \pi ^2}{27}+\frac{8 \gamma ^3}{9}+\frac{8 \gamma
     ^2}{9}-\frac{173}{72}\Bigg].
\end{align}
\endgroup
The Sudakov factor is
\begin{align} \label{Dsud}
&\ln D^{sud}_{Q\to Q}(N,\muzf,m) = \int_0^1 dz \frac{z^{N-1}-1}{1-z}\left[\int^{\muzf^2}_{m^2(1-z)^2} \frac{d\mu^2}{\mu^2}\, A[\as^{(n_l)}(\mu)] + H[\as^{(n_l)}(m(1-z))]\right]  \nonumber \\
&= \ln N g_1^{ini}(\lambda_0)+g_2^{ini}(\lambda_0;\muzf,\muzr,m)+\as^{(n_l)}(\muzr) g_3^{ini}(\lambda_0;\muzf,\muzr,m) + {\cal O}(\as^k \ln^{k-2}N)\, ,
\end{align}
where we have defined $\lambda_0=b_0^{(n_l)}\as^{(n_l)}(\muzr)\ln N$, and used
\be
\label{eq:binl}
b_i=b_i^{(n_l)} \quad \text{for}\quad  i\in\{0,1,2\}, \quad \quad  A_j=A_j^{(n_l)} \quad \text{for}\quad j\in\{2,3\},\quad \quad H_2 = H_2^{(n_l)}
\ee
we can write explicitly the $g_k^{ini}$ functions (that we have independently verified) as~\cite{Cacciari:2001cw,Catani:2003zt,Maltoni:2022bpy} 
\begingroup
\allowdisplaybreaks
\begin{align} \label{g1g2ini}
    g_1^{ini}(\lambda_0)&= - \frac{A_1}{2\pi b_0\lambda_0}[2\lambda_0+(1-2\lambda_0)\ln(1-2\lambda_0)] \, , \\
    g_2^{ini}(\lambda_0;\muzf,\muzr,m)&=\frac{A_1}{2\pi b_0}\left(\ln\frac{\muzf^2}{m^2}+2\gamma_E\right)\ln(1-2\lambda_0)\nonumber\\
    -&\frac{A_1 b_1}{2\pi b_0^3}(2\lambda_0+\ln(1-2\lambda_0)+\frac{1}{2}\ln^2(1-2\lambda_0))\nonumber\\
    +&\frac{1}{2\pi b_0}(2\lambda_0+\ln(1-2\lambda_0))\left(\frac{A_2 }{\pi b_0}+A_1\ln\frac{\muzr^2}{\muzf^2}\right)+\frac{H_1 }{2\pi b_0}\ln(1-2\lambda_0) \, , \\
        g_3^{ini}(\lambda_0;\muzf,\muzr,m)&=\frac{2A_1}{\pi}(\gamma_E^2+\zeta(2))\frac{\lambda_0}{1-2\lambda_0}+\frac{A_1\gamma_E b_1}{\pi b_0^2(1-2\lambda_0)}(2\lambda_0+\ln(1-2\lambda_0))\nonumber\\
    -&\frac{A_1b_1^2}{2\pi b_0^4(1-2\lambda_0)}\left(2\lambda_0^2+2\lambda_0\ln(1-2\lambda_0)+\frac{1}{2}\ln^2(1-2\lambda_0)\right)\nonumber\\
    -&\frac{A_1 b_2}{2\pi b_0^3}\left(2\lambda_0+\ln(1-2\lambda_0)+\frac{2\lambda_0^2}{1-2\lambda_0}\right)-\frac{A_3 }{\pi^3b_0^2}\frac{\lambda_0^2}{1-2\lambda_0}\nonumber\\
    +&\frac{H_1 b_1}{2\pi b_0^2}\left(\frac{2\lambda_0+\ln(1-2\lambda_0)}{1-2\lambda_0}\right)-\frac{H_1 \gamma_E}{\pi}\frac{2\lambda_0}{1-2\lambda_0}-\frac{H_2 }{\pi^2 b_0}\frac{\lambda_0}{1-2\lambda_0} \nonumber\\
    -&\frac{2\gamma_E A_2 }{\pi^2 b_0}\frac{\lambda_0}{1-2\lambda_0}+\frac{A_2 b_1}{2\pi^2b_0^3}\frac{1}{1-2\lambda_0}(2\lambda_0+\ln(1-2\lambda_0)+2\lambda_0^2) \nonumber \\
    +&\frac{A_2 }{\pi^2 b_0}\lambda_0 \ln\frac{m^2}{\muzf^2}+\frac{A_1}{2\pi}\lambda_0\ln^2\frac{m^2}{\muzf^2}-\frac{A_1}{\pi}\lambda_0\ln\frac{m^2}{\muzr^2}\ln\frac{m^2}{\muzf^2} \nonumber \\
    -&\frac{1}{1-2\lambda_0}\left[\frac{A_1 b_1}{2\pi b_0^2}(2\lambda_0+\ln(1-2\lambda_0))-\frac{2A_1\gamma_E}{\pi}\lambda_0-\frac{2A_2 }{\pi^2b_0}\lambda_0^2\right]\ln\frac{m^2}{\muzr^2}\nonumber \\
    -&\frac{A_1}{\pi}\frac{\lambda_0^2}{1-2\lambda_0}\ln^2\frac{m^2}{\muzr^2}+\frac{H_1 }{\pi}\frac{\lambda_0}{1-2\lambda_0}\ln\frac{m^2}{\muzr^2} \, .
\end{align}
\endgroup

For completeness, the expansion of the Sudakov factor up to $\mcO\left(\as^2\right)$ is
\begin{align}
[D^{sud}_{Q\to Q}(\muzf,m)]_{\as^2} &=1+\frac{\aS^{(n_l)}(\muzr)}{2\pi}\left.D^{sud}_{Q\to Q}(N,\muzf,m) \right|_{\as}\nonumber \\
&+ \left(\frac{\aS^{(n_l)}(\muzr)}{2\pi}\right)^2 \left.D^{sud}_{Q\to Q}(N,\muzf,\muzr,m)\right|_{\as^2}  \, ,
\end{align}
with coefficients
\begingroup
\allowdisplaybreaks
\begin{align}
&\left.D^{sud}_{Q\to Q}(N,\muzf,m)\right|_{\as}=-2\left[H_1 +A_1\left(2\gamma_E-\ln\frac{m^2}{\muzf^2}\right)\right]\ln N-2A_1\ln^2 N  \, , \\
&\left.D^{sud}_{Q\to Q}(N,\muzf,\muzr,m)\right|_{\as^2}=-4 \bigg(2A_2 \gamma_E+H_2 +2H_1 b_0\gamma_E\pi+2A_1 b_0\pi\left(\gamma_E^2+\zeta(2)\right)\nonumber \\
&~~~~~~~-2A_1 b_0\gamma_E\pi\ln\frac{m^2}{\muzr^2}-H_1 b_0\pi\ln\frac{m^2}{\muzr^2}-A_2 \ln\frac{m^2}{\muzf^2}+A_1 b_0\pi\ln\frac{m^2}{\muzr^2}\ln\frac{m^2}{\muzf^2}\nonumber \\
&~~~~~~~-\frac{1}{2}A_1 b_0\pi\ln^2\frac{m^2}{\muzf^2}\bigg)\ln N -4\pi\left(\frac{A_2 }{\pi}+2A_1 b_0\gamma_E+H_1 b_0-A_1 b_0\ln\frac{m^2}{\muzr^2}\right)\ln^2 N \nonumber \\
&~~~~~~~-\frac{8\pi}{3}A_1 b_0\ln^3N+2\left(-2A_1\gamma_E\ln N - H_1 \ln N+A_1\ln\frac{m^2}{\muzf^2}\ln N-A_1\ln^2N\right)^2 \,  ,
\end{align} 
\endgroup
having used again eq.~(\ref{eq:binl}).

\section{Landau pole prescription}
\label{app:landau}

As a manifestation of non-perturbative phenomena and eventual failure of perturbation theory, the Sudakov factors of initial condition eq.~\eqref{Dsud} and coefficient function eq.~\eqref{Csud} present singularities at large-$N$ values. The latter manifest for specific values of the parameters $\lambda$ and $\lambda_0$ at the so called Landau poles, sensitive to the values of $\as$.

For the coefficient function the Landau pole $N^L$ is at
\begin{equation}
\lambda = 1 \rightarrow N^L = \exp \left(\frac{1}{b_0 \aS(\mur)}\right) \,  ,
\end{equation}
and for the values of $\aS$ used appears at very large values of $N$ ($>100$), having no impact on the perturbative prediction.  For the initial conditions instead the Landau pole  $N_0^L$ is at
\begin{equation}
\lambda_0 = \frac{1}{2} \rightarrow N_{0}^L = \exp \left(\frac{1}{2b_0 \aS(\muzr)}\right) \,  ,
\end{equation}
and with typical values of $\as$ and $n_f$ we find that for the bottom $N_{0}^L\sim 30$ while for the charm $N_{0}^L\sim 7$ as can be seen in figures~\ref{fig:bottom-ini} and~\ref{fig:charm-ini-ratios}.  The behaviour of the cross section $\sigma_Q(N)$ for  production of a heavy quark in the large-$N$ region is dominated by the Landau pole of the initial condition and, for $N>N_0^L$, the moments can become negative and unphysical.
In order to produce sensible phenomenological results, a prescription must be applied to regulate the pole, and many choices can be made. 
We notice that a reasonable Landau prescription should not introduce power corrections that are larger than generically expected for the process in question~\cite{Cacciari:2005ry}.

A prescription that respects this condition is the so called CNO prescription (Cacciari-Nason-Oleari), that regulates the pole via a tunable parameter $f$.  It amounts to the following remapping of $N$ in the initial conditions 
\begin{equation}
N\to N\frac{1+f/N_{0}^L}{1+f N/N_{0}^L} \, ,
\end{equation}
and analogously in the coefficient functions replacing $N_0^L$ with $N^L$. The Landau singularity is regulated when $f\geq 1$ and we have chosen $f=1.25$ for the bottom and $f=1.25$ or $f=2.0$ for the charm. 

In general it is ambiguous whether one should apply such a prescription also to the expansion of the Sudakov factors when matching to the fixed order, since it is free from the Landau singularity.  Both choices are acceptable and regulate successfully the Landau pole, but give different results.  As in section \ref{sec:ini}, we denote as CNO the prescription in which the re-mapping is applied to the Sudakov factor and its expansion, and with CNOmod the prescription in which the remapping is applied only to the Sudakov factor. Since the resummed result and its expansion are both rescaled and cancel well, the CNO choice ensures that the fixed-order result is well preserved at small $N$. On the other hand, at large $N$ the (unrescaled) fixed-order result and the (rescaled) expansion of the resummation will not cancel perfectly. While this could be seen as spoiling the resummation, it mainly happens around and beyond the Landau pole, where phenomenology is anyway expected to be mainly driven by empirical modelling of non-perturbative effects. CNOmod ensures the large-$N$ cancellation, but it does not preserve the fixed-order calculation at small $N$.
Finally, we note that a third choice may in principle be legitimate, namely rescaling all three components (fixed-order, resummed, and expansion) of the matched result. We refrain from considering this option, which has the same properties of CNOmod.

In ref.~\cite{Czakon:2022pyz} the authors introduce a new prescription for the Landau pole that we have called CGMP (Czakon-Generet-Mitov-Poncelet). The exponents of the Sudakov factors in eqs.~(\ref{Csud}) and (\ref{Dsud}) are expanded and truncated in such a way that the desired logarithmic accuracy is preserved.  This amounts to the following truncation:
\begin{align}
 \ln N &g_1(\lambda)+g_2(\lambda)+\as g_3(\lambda)\nonumber \\
 &\simeq g_{1,2}\as\ln^2(N)+g_{1,3}\as^2\ln^3(N)+g_{1,4}\as^3\ln^4(N)+g_{1,5}\as^4\ln^5(N)+g_{1,6}\as^5\ln^6(N) \nonumber \\
 &+g_{2,1}\as\ln(N)+g_{2,2}\as^2\ln^2(N)+g_{2,3}\as^3\ln^3(N)+g_{2,4}\as^4\ln^4(N) \nonumber \\
 &+g_{3,1}\as^2\ln(N)+g_{3,2}\as^3\ln^2(N) \, .
\end{align}
The authors show that this prescription preserves the NNLL accuracy. They also show that it introduces power corrections in the initial condition that are larger than those expected from non-perturbative effects, but argue that such corrections are numerically small for the bottom quark case and that they are even smaller than those introduced by the CNO remapping.  

As mentioned in section \ref{sec:ini}, we have decided to adopt the CNO prescription for our comparisons with the data since the CGMP prescription yields poorly behaving results for the charm.  This can be seen both in $N$-space, where the CGMP-regularised cross section becomes smaller than the data at relatively small $N$-values, and in $x$-space, where the perturbative cross section has a steep fall to zero around $x\sim 0.8$ (see figure~\ref{fig:pertC}), making it impossible to describe the data with any of the simple non perturbative prescriptions used in~\cite{Cacciari:2005uk}\footnote{All non-perturbative functional forms are normalised functions that can only lower the value of the moments of the perturbative predictions.}.

\section{MELA Setup}
\label{app:melasetup}

In the following we detail the choices made in the usage of the evolution library MELA~\cite{Bertone:2015cwa}. 

The obvious choice for our process is the evolution of type `TIME', setting to `OFF' the polarization mode. 
We fix the method used in the DGLAP solution to `TRN' and the quark masses to be of `POLE' type.  We observe negligible discrepancies under the choice of the DGLAP solution method.  Our inputs for the values of quarks masses and flavour thresholds are given in section~\ref{sec:numres}.

For the value of $\as$ at a given scale in the initial conditions and coefficient functions we have always used the internal routine of MELA for $\as$ in a VFNS starting from a reference value of $\as(91.2 ~\mathrm{GeV}) = 0.118$. Where needed, we apply a correction at the threshold for the scheme used in the calculation of the fixed order ingredients.

For the evolution, the `VFNS' mode is always our default.  Given that the matching conditions at thresholds~\cite{Cacciari:2005ry} are not available in the time-like case at NNLO accuracy, the VFNS of MELA can normally only be used with a NLO evolution. With a small modification in the source code we have allowed for a NNLO evolution with only NLO matching thresholds. Such modification is now available in the branch \url{https://github.com/vbertone/MELA/tree/timelike_thresholds}. The numerical impact of the unknown and missing NNLO matching thresholds is expected to be negligible.\footnote{This can be conjectured observing that the effect of switching on and off in MELA the existing NLO matching thresholds is already negligible. Of course, only an explicit calculation can tell for sure.}

\section{Additional Numerical results}

\subsection{Initial conditions}
\label{app:ini}

\begin{figure}[t]
\includegraphics[width=\textwidth]{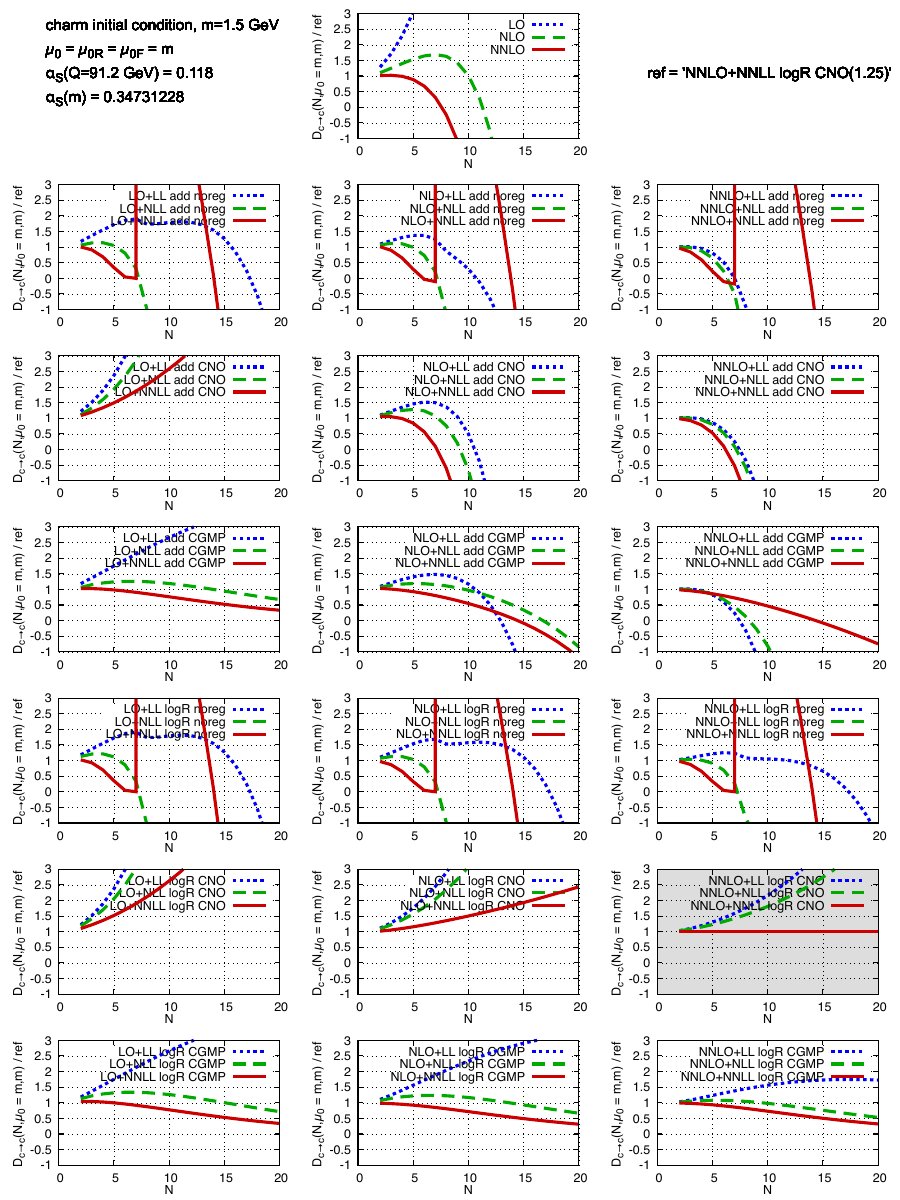}
\caption{\label{fig:charm-ini-ratios}The initial condition for the charm-to-charm fragmentation function, without any evolution, normalised to the curve `NNLO+NNLL logR CNO'. `CNO' is always used here with the parameter $f=1.25$.}
\end{figure}

\begin{figure}[t]
\includegraphics[width=\textwidth]{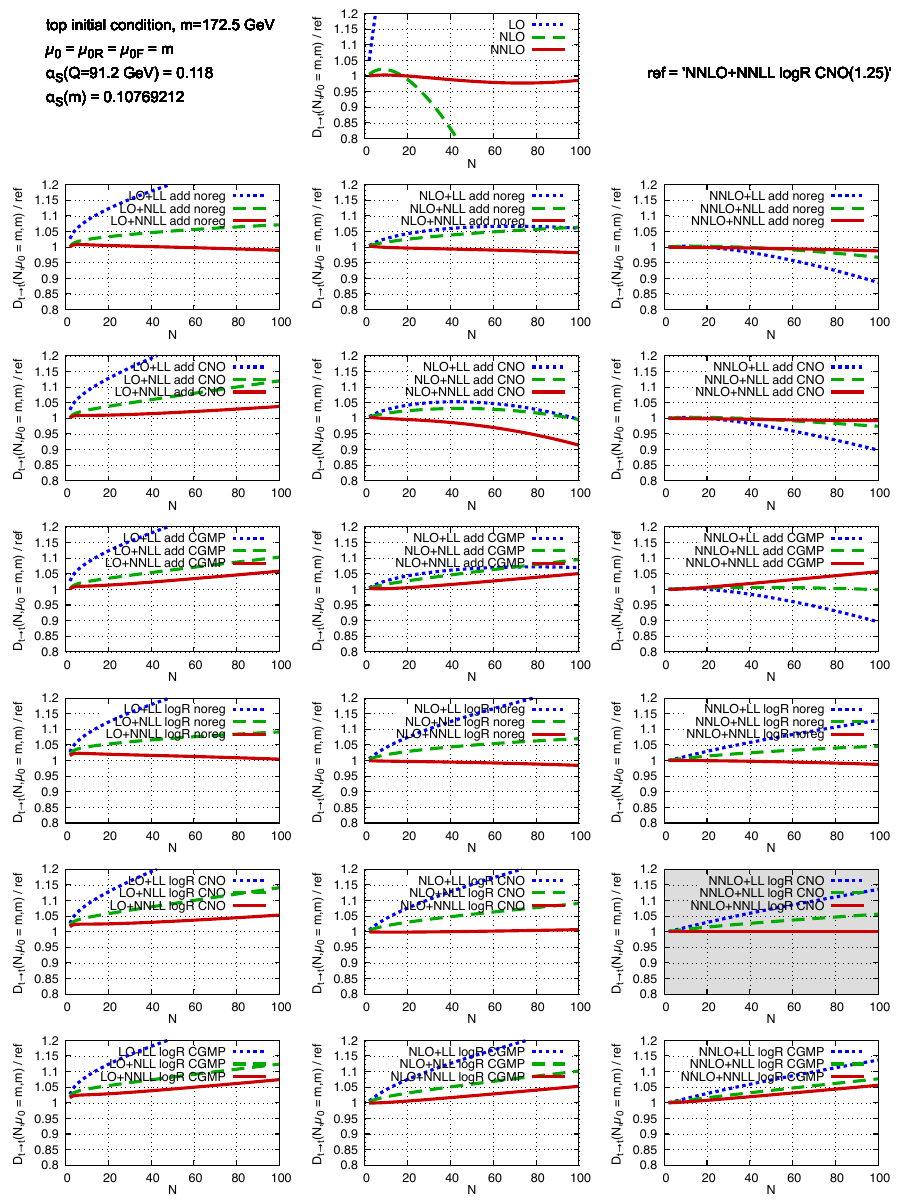}
\caption{\label{fig:top-ini-ratios}The initial condition for the top-to-top fragmentation function, without any evolution, normalised to the curve `NNLO+NNLL logR CNO'. `CNO' is always used here with the parameter $f=1.25$.}
\end{figure}

For completeness we show here, in figures~\ref{fig:charm-ini-ratios} and \ref{fig:top-ini-ratios}, the analogous of figure~\ref{fig:bottom-ini-ratios} for charm and top quarks. The pattern of convergence, or lack thereof, of the perturbative series and of the resummation is consistent with the bottom quark case, though respectively more and less extreme for charm and top quarks respectively. This is due to the relevant scale -- equal to the quark mass -- being smaller in the charm case and larger in the top case, and the corresponding strong coupling consequently larger and smaller.

\clearpage

\subsection[Full $e^+e^-$ fragmentation function]{Full \boldmath{$e^+e^-$} fragmentation function}
\label{app:fullff}

\begin{figure}[t]
\includegraphics[width=\textwidth]{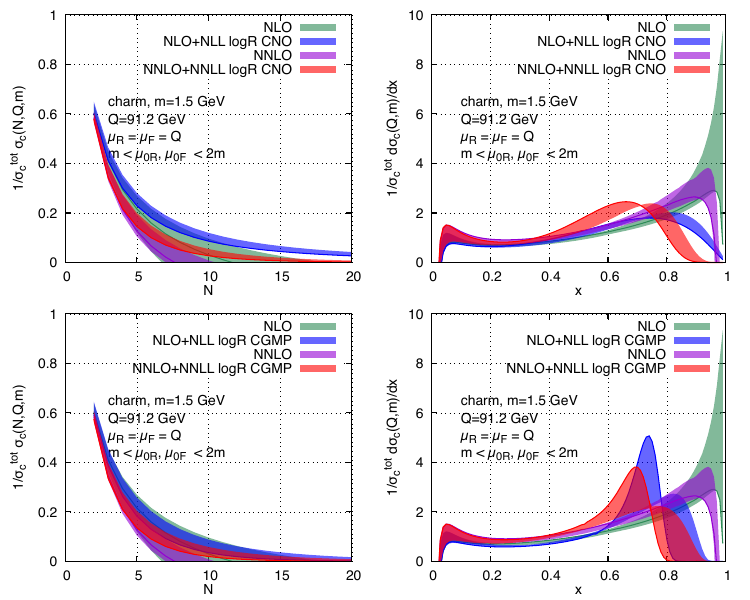}
\caption{\label{fig:ff-charm-ini} Charm fragmentation function. Uncertainty bands due to variation of the $\muzr$ and $\muzf$ scales around $m$. Left plots in $N$-space, right plots in $x$-space. Upper plots with CNO regularisation ($f=1.25$), lower plots with CGMP regularisation. Note that, at variance with the bottom and top quark cases, here we only vary the renormalisation scale in the range $m <\muzr<2m$.}
\end{figure}

\begin{figure}[t]
\includegraphics[width=\textwidth]{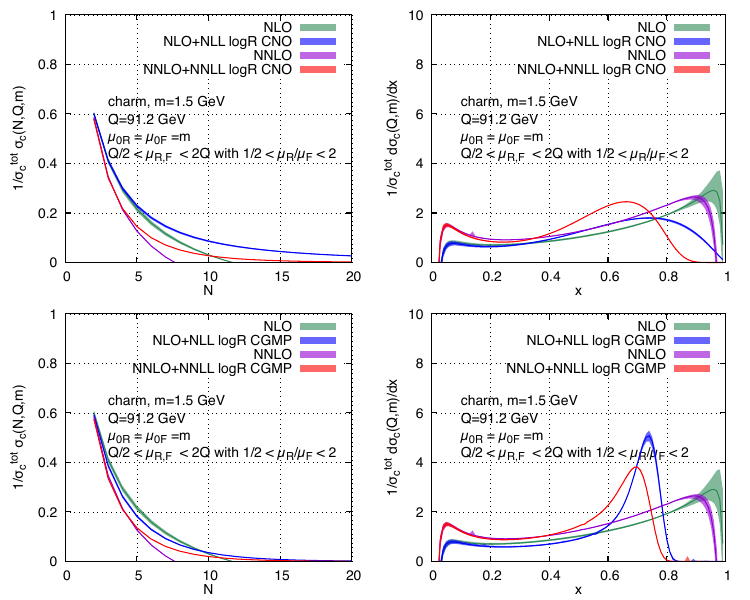}
\caption{\label{fig:ff-charm} Charm fragmentation function. Uncertainty bands due to variation of the $\mur$ and $\muf$ scales around $Q$. Left plots in $N$-space, right plots in $x$-space. Upper plots with CNO regularisation ($f=1.25$), lower plots with CGMP regularisation
The little spikes that one can observe on the curves in $x$ space, here and possibly elsewhere, are artefacts of the numerical Mellin inversion.}
\end{figure}

Again for completeness, we show here the scale dependence of the full $e^+e^-$ fragmentation function into charm (at $Q=91.2$~GeV) and into top (at $Q=1000$~GeV) quarks. 
Note that, at variance with the bottom and top quark cases, for the charm quark we only vary the renormalisation scale in the range $m <\muzr<2m$, to avoid reaching too deep into a barely perturbative region.
These figures are to be compared with the ones for bottom quarks, in figures~\ref{fig:ff-bottom-ini} and~\ref{fig:ff-bottom}.

One can see that in the charm quark case (figures~\ref{fig:ff-charm-ini} and~\ref{fig:ff-charm}) the uncertainty bands are wider than in the bottom quark case. In the case of scale variations around the initial scale $m$, and despite the reduced range of variation of the renormalisation scale, any perturbative hierarchy is essentially absent, and the bands are so large that they make perturbative predictivity quite weak.  Sensitivity on the choice of regularisation of the Landau pole, clearly seen in the $x$-space plots, also appears to be even larger than in the bottom case.

\begin{figure}[t]
\includegraphics[width=\textwidth]{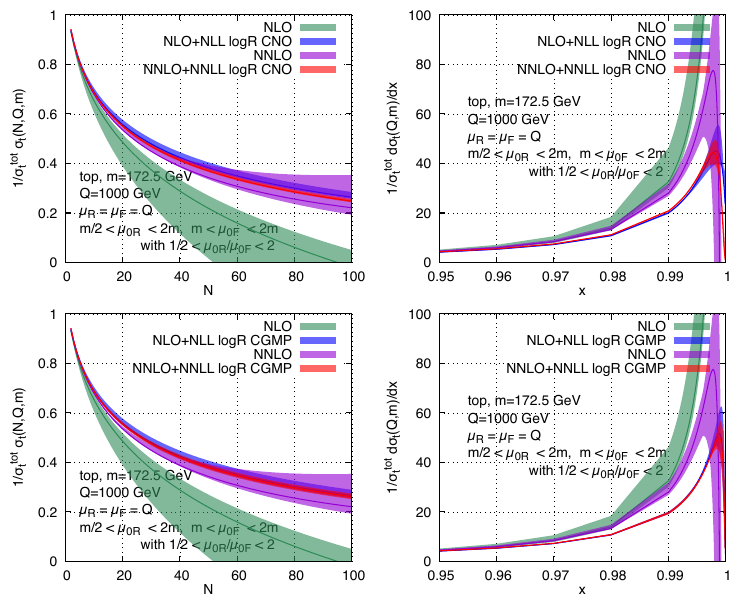}
\caption{\label{fig:ff-top-ini} Top fragmentation function at $Q=1000$~GeV. Uncertainty bands due to variation of the $\muzr$ and $\muzf$ scales around $m$. Left plots in $N$-space, right plots in $x$-space. Upper plots with CNO regularisation ($f=1.25$), lower plots with CGMP regularisation. Note that the ranges of both the horizontal and the vertical axis in the $x$-space plots differ from those used for charm and for bottom.}
\end{figure}

For the top quark case in figure~\ref{fig:ff-top-ini} instead, thanks to the small value of $\as$ at all scales involved in the process ($m=172.5$~GeV and $Q=1000$~GeV), the perturbative behaviour is much better. The bands in $N$-space display the expected hierarchy, and in $x$-space all curves are very similar, except very close to $x=1$, which is the only region shown in the plot. While the top case is presently not of phenomenological interest, it serves as a cross-check that the `confusing' behaviour observed in the bottom quark and -- to an even larger extent -- in the charm quark case can be ascribed to the large value of $\as$ in play there. We do not show results for variations of the final scales around $Q$ for the top case as they result in bands even narrower than in the bottom case and are of no pedagogical value.

\clearpage
\addcontentsline{toc}{section}{References}
\bibliographystyle{JHEP}
\bibliography{paper}

\providecommand{\href}[2]{#2}\begingroup\raggedright\begin{thebibliography}{10}

\bibitem{Suzuki:1977km}
M.~Suzuki, \emph{{Fragmentation of Hadrons from Heavy Quark Partons}},
  \href{https://doi.org/10.1016/0370-2693(77)90761-4}{\emph{Phys. Lett. B}
  {\bfseries 71} (1977) 139}.

\bibitem{Bjorken:1977md}
J.D.~Bjorken, \emph{{Properties of Hadron Distributions in Reactions Containing
  Very Heavy Quarks}},
  \href{https://doi.org/10.1103/PhysRevD.17.171}{\emph{Phys. Rev. D} {\bfseries
  17} (1978) 171}.

\bibitem{Kinoshita:1981af}
K.~Kinoshita, \emph{{Heavy Quark Fragmentation}},  4, 1981.

\bibitem{Kinoshita:1985mh}
K.~Kinoshita, \emph{{A Covariant Parton Model for Heavy Quark Fragmentation}},
  \href{https://doi.org/10.1143/PTP.75.84}{\emph{Prog. Theor. Phys.} {\bfseries
  75} (1986) 84}.

\bibitem{Peterson:1982ak}
C.~Peterson, D.~Schlatter, I.~Schmitt and P.M.~Zerwas, \emph{{Scaling
  Violations in Inclusive $e^+e^-$ Annihilation Spectra}},
  \href{https://doi.org/10.1103/PhysRevD.27.105}{\emph{Phys. Rev. D} {\bfseries
  27} (1983) 105}.

\bibitem{Mele:1990yq}
B.~Mele and P.~Nason, \emph{{Next-to-leading QCD calculation of the heavy quark
  fragmentation function}},
  \href{https://doi.org/10.1016/0370-2693(90)90704-A}{\emph{Phys. Lett. B}
  {\bfseries 245} (1990) 635}.

\bibitem{Mele:1990cw}
B.~Mele and P.~Nason, \emph{{The Fragmentation function for heavy quarks in
  QCD}}, \href{https://doi.org/10.1016/0550-3213(91)90597-Q}{\emph{Nucl. Phys.
  B} {\bfseries 361} (1991) 626}.

\bibitem{Dokshitzer:1995ev}
Y.L.~Dokshitzer, V.A.~Khoze and S.I.~Troian, \emph{{Specific features of heavy
  quark production. LPHD approach to heavy particle spectra}},
  \href{https://doi.org/10.1103/PhysRevD.53.89}{\emph{Phys. Rev. D} {\bfseries
  53} (1996) 89} [\href{https://arxiv.org/abs/hep-ph/9506425}{{\ttfamily
  hep-ph/9506425}}].

\bibitem{Cacciari:2001cw}
M.~Cacciari and S.~Catani, \emph{{Soft gluon resummation for the fragmentation
  of light and heavy quarks at large x}},
  \href{https://doi.org/10.1016/S0550-3213(01)00469-2}{\emph{Nucl. Phys. B}
  {\bfseries 617} (2001) 253}
  [\href{https://arxiv.org/abs/hep-ph/0107138}{{\ttfamily hep-ph/0107138}}].

\bibitem{Cacciari:2005uk}
M.~Cacciari, P.~Nason and C.~Oleari, \emph{{A Study of heavy flavored meson
  fragmentation functions in $e^+e^-$ annihilation}},
  \href{https://doi.org/10.1088/1126-6708/2006/04/006}{\emph{JHEP} {\bfseries
  04} (2006) 006} [\href{https://arxiv.org/abs/hep-ph/0510032}{{\ttfamily
  hep-ph/0510032}}].

\bibitem{Aglietti:2007bp}
U.~Aglietti, L.~Di~Giustino, G.~Ferrera, A.~Renzaglia, G.~Ricciardi and
  L.~Trentadue, \emph{{Threshold Resummation in $B \to X_c\,l\,\nu_l$ Decays}},
  \href{https://doi.org/10.1016/j.physletb.2007.07.041}{\emph{Phys. Lett. B}
  {\bfseries 653} (2007) 38} [\href{https://arxiv.org/abs/0707.2010}{{\ttfamily
  0707.2010}}].

\bibitem{Aglietti:2022rcm}
U.G.~Aglietti and G.~Ferrera, \emph{{Improved factorization for threshold
  resummation in heavy quark to heavy quark decays}},
  \href{https://doi.org/10.1140/epjc/s10052-023-11440-y}{\emph{Eur. Phys. J. C}
  {\bfseries 83} (2023) 335}
  [\href{https://arxiv.org/abs/2211.14397}{{\ttfamily 2211.14397}}].

\bibitem{Gaggero:2022hmv}
D.~Gaggero, A.~Ghira, S.~Marzani and G.~Ridolfi, \emph{{Soft logarithms in
  processes with heavy quarks}},
  \href{https://doi.org/10.1007/JHEP09(2022)058}{\emph{JHEP} {\bfseries 09}
  (2022) 058} [\href{https://arxiv.org/abs/2207.13567}{{\ttfamily
  2207.13567}}].

\bibitem{Ghira:2023bxr}
A.~Ghira, S.~Marzani and G.~Ridolfi, \emph{{A consistent resummation of mass
  and soft logarithms in processes with heavy flavours}},
  \href{https://doi.org/10.1007/JHEP11(2023)120}{\emph{JHEP} {\bfseries 11}
  (2023) 120} [\href{https://arxiv.org/abs/2309.06139}{{\ttfamily
  2309.06139}}].

\bibitem{Jaffe:1993ie}
R.L.~Jaffe and L.~Randall, \emph{{Heavy quark fragmentation into heavy
  mesons}}, \href{https://doi.org/10.1016/0550-3213(94)90495-2}{\emph{Nucl.
  Phys. B} {\bfseries 412} (1994) 79}
  [\href{https://arxiv.org/abs/hep-ph/9306201}{{\ttfamily hep-ph/9306201}}].

\bibitem{Neubert:2007je}
M.~Neubert, \emph{{Factorization analysis for the fragmentation functions of
  hadrons containing a heavy quark}},
  \href{https://arxiv.org/abs/0706.2136}{{\ttfamily 0706.2136}}.

\bibitem{Gardi:2003ar}
E.~Gardi and M.~Cacciari, \emph{{Perturbative and nonperturbative aspects of
  heavy quark fragmentation}},
  \href{https://doi.org/10.1140/epjcd/s2003-03-1002-y}{\emph{Eur. Phys. J. C}
  {\bfseries 33} (2004) S876}
  [\href{https://arxiv.org/abs/hep-ph/0308235}{{\ttfamily hep-ph/0308235}}].

\bibitem{Gardi:2005yi}
E.~Gardi, \emph{{On the quark distribution in an on-shell heavy quark and its
  all-order relations with the perturbative fragmentation function}},
  \href{https://doi.org/10.1088/1126-6708/2005/02/053}{\emph{JHEP} {\bfseries
  02} (2005) 053} [\href{https://arxiv.org/abs/hep-ph/0501257}{{\ttfamily
  hep-ph/0501257}}].

\bibitem{Aglietti:2006yf}
U.~Aglietti, G.~Corcella and G.~Ferrera, \emph{{Modelling non-perturbative
  corrections to bottom-quark fragmentation}},
  \href{https://doi.org/10.1016/j.nuclphysb.2007.04.014}{\emph{Nucl. Phys. B}
  {\bfseries 775} (2007) 162}
  [\href{https://arxiv.org/abs/hep-ph/0610035}{{\ttfamily hep-ph/0610035}}].

\bibitem{Corcella:2007tg}
G.~Corcella and G.~Ferrera, \emph{{Charm-quark fragmentation with an effective
  coupling constant}},
  \href{https://doi.org/10.1088/1126-6708/2007/12/029}{\emph{JHEP} {\bfseries
  12} (2007) 029} [\href{https://arxiv.org/abs/0706.2357}{{\ttfamily
  0706.2357}}].

\bibitem{Fickinger:2016rfd}
M.~Fickinger, S.~Fleming, C.~Kim and E.~Mereghetti, \emph{{Effective field
  theory approach to heavy quark fragmentation}},
  \href{https://doi.org/10.1007/JHEP11(2016)095}{\emph{JHEP} {\bfseries 11}
  (2016) 095} [\href{https://arxiv.org/abs/1606.07737}{{\ttfamily
  1606.07737}}].

\bibitem{Rijken:1996vr}
P.J.~Rijken and W.L.~van Neerven, \emph{{$\mathcal{O}(\alpha^2_s)$
  contributions to the longitudinal fragmentation function in $e^+e^-$
  annihilation}},
  \href{https://doi.org/10.1016/0370-2693(96)00898-2}{\emph{Phys. Lett. B}
  {\bfseries 386} (1996) 422}
  [\href{https://arxiv.org/abs/hep-ph/9604436}{{\ttfamily hep-ph/9604436}}].

\bibitem{Rijken:1996ns}
P.J.~Rijken and W.L.~van Neerven, \emph{{Higher order QCD corrections to the
  transverse and longitudinal fragmentation functions in electron - positron
  annihilation}},
  \href{https://doi.org/10.1016/S0550-3213(96)00669-4}{\emph{Nucl. Phys. B}
  {\bfseries 487} (1997) 233}
  [\href{https://arxiv.org/abs/hep-ph/9609377}{{\ttfamily hep-ph/9609377}}].

\bibitem{Mitov:2006wy}
A.~Mitov and S.-O.~Moch, \emph{{QCD Corrections to Semi-Inclusive Hadron
  Production in Electron-Positron Annihilation at Two Loops}},
  \href{https://doi.org/10.1016/j.nuclphysb.2006.05.018}{\emph{Nucl. Phys. B}
  {\bfseries 751} (2006) 18}
  [\href{https://arxiv.org/abs/hep-ph/0604160}{{\ttfamily hep-ph/0604160}}].

\bibitem{Blumlein:2006rr}
J.~Blumlein and V.~Ravindran, \emph{{$\mathcal{O}(\alpha^2_s)$ Timelike Wilson
  Coefficients for Parton-Fragmentation Functions in Mellin Space}},
  \href{https://doi.org/10.1016/j.nuclphysb.2006.04.032}{\emph{Nucl. Phys. B}
  {\bfseries 749} (2006) 1}
  [\href{https://arxiv.org/abs/hep-ph/0604019}{{\ttfamily hep-ph/0604019}}].

\bibitem{Melnikov:2004bm}
K.~Melnikov and A.~Mitov, \emph{{Perturbative heavy quark fragmentation
  function through $\mathcal{O}(\alpha^2_s)$}},
  \href{https://doi.org/10.1103/PhysRevD.70.034027}{\emph{Phys. Rev. D}
  {\bfseries 70} (2004) 034027}
  [\href{https://arxiv.org/abs/hep-ph/0404143}{{\ttfamily hep-ph/0404143}}].

\bibitem{Mitov:2004du}
A.~Mitov, \emph{{Perturbative heavy quark fragmentation function through
  $\mathcal{O}(\alpha^2_s)$: Gluon initiated contribution}},
  \href{https://doi.org/10.1103/PhysRevD.71.054021}{\emph{Phys. Rev. D}
  {\bfseries 71} (2005) 054021}
  [\href{https://arxiv.org/abs/hep-ph/0410205}{{\ttfamily hep-ph/0410205}}].

\bibitem{Ridolfi:2019bch}
G.~Ridolfi, M.~Ubiali and M.~Zaro, \emph{{A fragmentation-based study of heavy
  quark production}},
  \href{https://doi.org/10.1007/JHEP01(2020)196}{\emph{JHEP} {\bfseries 01}
  (2020) 196} [\href{https://arxiv.org/abs/1911.01975}{{\ttfamily
  1911.01975}}].

\bibitem{Maltoni:2022bpy}
F.~Maltoni, G.~Ridolfi, M.~Ubiali and M.~Zaro, \emph{{Resummation effects in
  the bottom-quark fragmentation function}},
  \href{https://doi.org/10.1007/JHEP10(2022)027}{\emph{JHEP} {\bfseries 10}
  (2022) 027} [\href{https://arxiv.org/abs/2207.10038}{{\ttfamily
  2207.10038}}].

\bibitem{Czakon:2022pyz}
M.~Czakon, T.~Generet, A.~Mitov and R.~Poncelet, \emph{{NNLO B-fragmentation
  fits and their application to $ t\overline{t} $ production and decay at the
  LHC}}, \href{https://doi.org/10.1007/JHEP03(2023)251}{\emph{JHEP} {\bfseries
  03} (2023) 251} [\href{https://arxiv.org/abs/2210.06078}{{\ttfamily
  2210.06078}}].

\bibitem{Collins:1981uk}
J.C.~Collins and D.E.~Soper, \emph{{Back-To-Back Jets in QCD}},
  \href{https://doi.org/10.1016/0550-3213(81)90339-4}{\emph{Nucl. Phys. B}
  {\bfseries 193} (1981) 381}.

\bibitem{Collins:1981uw}
J.C.~Collins and D.E.~Soper, \emph{{Parton Distribution and Decay Functions}},
  \href{https://doi.org/10.1016/0550-3213(82)90021-9}{\emph{Nucl. Phys. B}
  {\bfseries 194} (1982) 445}.

\bibitem{Bertone:2015cwa}
V.~Bertone, S.~Carrazza and E.R.~Nocera, \emph{{Reference results for time-like
  evolution up to $\mathcal{O}\left({\alpha}_s^3\right) $}},
  \href{https://doi.org/10.1007/JHEP03(2015)046}{\emph{JHEP} {\bfseries 03}
  (2015) 046} [\href{https://arxiv.org/abs/1501.00494}{{\ttfamily
  1501.00494}}].

\bibitem{Altarelli:1977zs}
G.~Altarelli and G.~Parisi, \emph{{Asymptotic Freedom in Parton Language}},
  \href{https://doi.org/10.1016/0550-3213(77)90384-4}{\emph{Nucl. Phys. B}
  {\bfseries 126} (1977) 298}.

\bibitem{Curci:1980uw}
G.~Curci, W.~Furmanski and R.~Petronzio, \emph{{Evolution of Parton Densities
  Beyond Leading Order: The Nonsinglet Case}},
  \href{https://doi.org/10.1016/0550-3213(80)90003-6}{\emph{Nucl. Phys. B}
  {\bfseries 175} (1980) 27}.

\bibitem{Furmanski:1980cm}
W.~Furmanski and R.~Petronzio, \emph{{Singlet Parton Densities Beyond Leading
  Order}}, \href{https://doi.org/10.1016/0370-2693(80)90636-X}{\emph{Phys.
  Lett. B} {\bfseries 97} (1980) 437}.

\bibitem{Moch:2004pa}
S.~Moch, J.A.M.~Vermaseren and A.~Vogt, \emph{{The Three loop splitting
  functions in QCD: The Nonsinglet case}},
  \href{https://doi.org/10.1016/j.nuclphysb.2004.03.030}{\emph{Nucl. Phys. B}
  {\bfseries 688} (2004) 101}
  [\href{https://arxiv.org/abs/hep-ph/0403192}{{\ttfamily hep-ph/0403192}}].

\bibitem{Vogt:2004mw}
A.~Vogt, S.~Moch and J.A.M.~Vermaseren, \emph{{The Three-loop splitting
  functions in QCD: The Singlet case}},
  \href{https://doi.org/10.1016/j.nuclphysb.2004.04.024}{\emph{Nucl. Phys. B}
  {\bfseries 691} (2004) 129}
  [\href{https://arxiv.org/abs/hep-ph/0404111}{{\ttfamily hep-ph/0404111}}].

\bibitem{Blumlein:2006pj}
J.~Blumlein and V.~Ravindran, \emph{{QCD threshold corrections to Higgs decay
  and to hadroproduction in $l^+l^-$ annihilation}},
  \href{https://doi.org/10.1016/j.physletb.2006.07.029}{\emph{Phys. Lett. B}
  {\bfseries 640} (2006) 40}
  [\href{https://arxiv.org/abs/hep-ph/0605011}{{\ttfamily hep-ph/0605011}}].

\bibitem{Moch:2009my}
S.~Moch and A.~Vogt, \emph{{Higher-order threshold resummation for
  semi-inclusive $e^+e^-$ annihilation}},
  \href{https://doi.org/10.1016/j.physletb.2009.09.001}{\emph{Phys. Lett. B}
  {\bfseries 680} (2009) 239}
  [\href{https://arxiv.org/abs/0908.2746}{{\ttfamily 0908.2746}}].

\bibitem{Catani:1992ua}
S.~Catani, L.~Trentadue, G.~Turnock and B.R.~Webber, \emph{{Resummation of
  large logarithms in e+ e- event shape distributions}},
  \href{https://doi.org/10.1016/0550-3213(93)90271-P}{\emph{Nucl. Phys. B}
  {\bfseries 407} (1993) 3}.

\bibitem{ALEPH:2001pfo}
{\scshape ALEPH} collaboration, \emph{{Study of the fragmentation of b quarks
  into B mesons at the Z peak}},
  \href{https://doi.org/10.1016/S0370-2693(01)00690-6}{\emph{Phys. Lett. B}
  {\bfseries 512} (2001) 30}
  [\href{https://arxiv.org/abs/hep-ex/0106051}{{\ttfamily hep-ex/0106051}}].

\bibitem{OPAL:2002plk}
{\scshape OPAL} collaboration, \emph{{Inclusive analysis of the b quark
  fragmentation function in Z decays at LEP}},
  \href{https://doi.org/10.1140/epjc/s2003-01229-x}{\emph{Eur. Phys. J. C}
  {\bfseries 29} (2003) 463}
  [\href{https://arxiv.org/abs/hep-ex/0210031}{{\ttfamily hep-ex/0210031}}].

\bibitem{SLD:2002poq}
{\scshape SLD} collaboration, \emph{{Measurement of the b quark fragmentation
  function in Z0 decays}},
  \href{https://doi.org/10.1103/PhysRevD.66.079905}{\emph{Phys. Rev. D}
  {\bfseries 65} (2002) 092006}
  [\href{https://arxiv.org/abs/hep-ex/0202031}{{\ttfamily hep-ex/0202031}}].

\bibitem{Barker:2002iuq}
G.~Barker, E.~Ben-Haim, M.~Feindt, U.~Kerzel, P.~Roudeau, L.~Ramler et~al.,
  \emph{{A Study of the b-Quark Fragmentation Function with the DELPHI Detector
  at LEP I}}, .

\bibitem{ALEPH:1999syy}
{\scshape ALEPH} collaboration, \emph{{Study of charm production in Z decays}},
  \href{https://doi.org/10.1007/s100520000421}{\emph{Eur. Phys. J. C}
  {\bfseries 16} (2000) 597}
  [\href{https://arxiv.org/abs/hep-ex/9909032}{{\ttfamily hep-ex/9909032}}].

\bibitem{CLEO:2004enr}
{\scshape CLEO} collaboration, \emph{{Charm meson spectra in $e^{+} e^{-}$
  annihilation at 10.5-GeV c.m.e.}},
  \href{https://doi.org/10.1103/PhysRevD.70.112001}{\emph{Phys. Rev. D}
  {\bfseries 70} (2004) 112001}
  [\href{https://arxiv.org/abs/hep-ex/0402040}{{\ttfamily hep-ex/0402040}}].

\bibitem{Belle:2005mtx}
{\scshape Belle} collaboration, \emph{{Charm hadrons from fragmentation and B
  decays in $e^+e^-$ annihilation at $\sqrt{s} = 10.6$~GeV}},
  \href{https://doi.org/10.1103/PhysRevD.73.032002}{\emph{Phys. Rev. D}
  {\bfseries 73} (2006) 032002}
  [\href{https://arxiv.org/abs/hep-ex/0506068}{{\ttfamily hep-ex/0506068}}].

\bibitem{Kartvelishvili:1977pi}
V.G.~Kartvelishvili, A.K.~Likhoded and V.A.~Petrov, \emph{{On the Fragmentation
  Functions of Heavy Quarks Into Hadrons}},
  \href{https://doi.org/10.1016/0370-2693(78)90653-6}{\emph{Phys. Lett. B}
  {\bfseries 78} (1978) 615}.

\bibitem{Nason:1997nu}
P.~Nason and C.~Oleari, \emph{{On the fragmentation function for heavy quarks
  in $e^+e^-$ collisions}},
  \href{https://doi.org/10.1016/S0370-2693(97)01414-7}{\emph{Phys. Lett. B}
  {\bfseries 418} (1998) 199}
  [\href{https://arxiv.org/abs/hep-ph/9709358}{{\ttfamily hep-ph/9709358}}].

\bibitem{Catani:2003zt}
S.~Catani, D.~de~Florian, M.~Grazzini and P.~Nason, \emph{{Soft gluon
  resummation for Higgs boson production at hadron colliders}},
  \href{https://doi.org/10.1088/1126-6708/2003/07/028}{\emph{JHEP} {\bfseries
  07} (2003) 028} [\href{https://arxiv.org/abs/hep-ph/0306211}{{\ttfamily
  hep-ph/0306211}}].

\bibitem{Cacciari:2005ry}
M.~Cacciari, P.~Nason and C.~Oleari, \emph{{Crossing heavy-flavor thresholds in
  fragmentation functions}},
  \href{https://doi.org/10.1088/1126-6708/2005/10/034}{\emph{JHEP} {\bfseries
  10} (2005) 034} [\href{https://arxiv.org/abs/hep-ph/0504192}{{\ttfamily
  hep-ph/0504192}}].

\end{thebibliography}\endgroup

\end{document}